\documentclass[10pt,tightenlines,eqsecnum,floats,aps,nofootinbib,prd]{revtex4-2}
\usepackage{amsmath,amssymb,amsfonts,amsthm,amscd}
\usepackage{graphicx}
\usepackage{amsmath}
\usepackage{amssymb}
\usepackage{graphicx}
\usepackage{cancel}
\usepackage{slashed}
\usepackage{tabularx}
\usepackage{mathtools}
\usepackage{tcolorbox}
\usepackage{tikz-cd}
\usepackage{tikz}
\usepackage{mathrsfs}
\usepackage{relsize}
\usepackage{natbib}
\usepackage[margin=1in]{geometry}
\usepackage{hyperref}
\usepackage[normalem]{ulem}
\usetikzlibrary{shapes.geometric, arrows}
\tikzstyle{startstop} = [rectangle, rounded corners, minimum width=3cm, minimum height=1cm,text centered, draw=black, fill=red!30]
\tikzstyle{arrow} = [thick,->,>=stealth]
\begin{document}
\title{Gauge Symmetries, Contact Reduction, and Singular Field Theories}
\author{Callum Bell}
\email{c.bell8@lancaster.ac.uk}
\affiliation{Department of Physics, Lancaster University, Lancaster UK}

\author{David Sloan}
\email{d.sloan@lancaster.ac.uk}
\affiliation{Department of Physics, Lancaster University, Lancaster UK}

\begin{abstract}
    The symmetry reduction of dynamical systems that are invariant under changes of global scale is well-understood for classical theories of particles, and fields. The excision of the superfluous degree of freedom generating such rescalings leads to a dynamically-equivalent theory, which is frictional in nature. In this article, we extend the formalism to physical models, of both particles and fields, described by singular Lagrangians. In order to work with a finite-dimensional (velocity) phase space, our construction requires that we treat classical field theories within the De-Donder Weyl formalism, in which a multisymplectic structure is introduced on the first jets of the bundle of fields. The results obtained are subsequently applied to a number of physically-motivated examples, as well as a discussion presented on the implications of our work for classical General Relativity.
\end{abstract}
\maketitle
\section{Introduction}
Singular field theories are often amongst the most interesting class of models encountered in the description of natural phenomena. Indeed, the most precisely tested mathematical theory to date, the Standard Model of Particle Physics, is a chiral quantum field theory, gauged under local transformations of the group $SU(3)_C\times SU(2)_L\times U(1)_Y$ \cite{peskin2018introduction}. Further, it is conjectured that in any UV-complete theory of quantum gravity, all symmetries must be either spontaneously broken, or gauged \cite{palti2019swampland}. It is thus clear that singular theories are of central importance to modern theoretical physics. In the present context, we take the word `singular' to refer to any theory whose Hamiltonian (or Lagrangian) description must, in some way, be constrained due to the presence of degeneracies.\\

It is often found that classical theories described within the Lagrangian or Hamiltonian framework are dependent upon a degree of freedom which is entirely inconsequential for the deduction of the dynamical evolution of the system's observables. In slightly more mathematical parlance, one may frequently identify a degree of freedom within the mathematical description of a system, without which the dynamical algebra of observables continues to close. It is trivially true that changes in a variable that may be excised from a dynamical algebra without affecting its closedness cannot be deduced from measurements of the remaining elements of the algebra. It follows, therefore, that there exists a description, entirely consistent with the dynamics of the elements of the algebra to which we have ontological access, which makes no reference to this degree of freedom.\\

As an example of this somewhat abstract line of reasoning, we might consider an isolated two-dimensional simple harmonic oscillator. The only dimensionful quantity in this system relating to length is the radial distance of the particle from the origin. Any observer internal to this system may perform measurements, but has access only to \textit{ratios} of dimensionful quantities. Indeed, since the only parameter with dimensions of length is the particle's radial displacement, it follows that the dynamical evolution of the oscillator requires knowledge of ratios of the radial displacements at different instants of time, but is wholly insensitive to any absolute value of this displacement. Stated equivalently, we could rescale the whole experimental configuration by a factor of, say, ten, and the subsequent dynamical evolution would remain unaltered. Referring to the discussion of the previous paragraph, we would say that the dynamical algebra of observables closes without reference to the value of radial displacement at a particular instant of time.\\

We propose that the correct interpretation of this situation is that the conventional description of the oscillator contains a redundant scaling degree of freedom, corresponding to the value of the radial displacement. Following the reasoning of Leibniz, a description that grants ontological distinction between two configurations which differ only by unobservable features should be rejected \cite{leibniz1989discourse}. The hallmark of the presence of such a redundant degree of freedom is invariance under a class of non-strictly canonical transformations known as scaling symmetries \cite{sloan2018dynamical,carinena2013canonoid}. Following the Leibnizian line of argumentation, we should seek to remove this redundant degree of freedom from our ontology. Indeed, this procedure - known as contact reduction - is well understood for regular theories of particles \cite{sloan2021scale} and fields \cite{bell2025dynamicalsimilaritymultisymplecticfield}. The objective of the present article is to construct a framework in which to analyse \textit{singular} theories that possess scaling symmetries.\\

For our study of theories of particles, we utilise the pre-symplectic constraint algorithm developed in \cite{Nester1}, and refer to \cite{bell2026constrained} for a somewhat more practical exposition. This same review also contains a detailed analysis of the geometry of the constraint surface, as well as a presentation of the analogous geometric constraint algorithm for pre-contact manifolds, and so we shall limit our discussion to a brief introduction to contact Hamiltonian systems. The reader familiar with contact geometry and its applications in the description of mechanical systems may therefore safely skip section (\ref{Sec:ContactGeom}).\\

Following a succinct discussion of the formalism of contact reduction for regular systems, the remainder of section (\ref{Sec:ContactReduction1}) contains novel ideas on how this process may be extended to cases in which degeneracies require constraints to be placed upon the dynamical system of interest. We also consider the commutative relationship between excising redundant scaling degrees of freedom, and the restriction of a system's phase space, finding that the order of operations is inconsequential for the final dynamics. Our formalism is then applied to a complete example.\\

The latter half of our work focuses on classical theories of fields. Both symplectic and contact geometry may be generalised to accommodate a finite-dimensional, manifestly covariant description of field theories, in which we work on the first jet bundle over a system's covariant configuration space. In \cite{bell2025dynamicalsimilaritymultisymplecticfield}, it was demonstrated that contact reduction may be extended to a field-theoretic context, and that once a scaling degree of freedom is eliminated, the multisymplectic structure with which we begin becomes multicontact \cite{roman2009multisymplectic,de2023multicontact}. Multicontact field theories have the peculiar feature that they are action-dependent; as a result, they describe systems which are non-conservative. Drawing upon the results of \cite{bell2025dynamicalsimilaritymultisymplecticfield}, section (\ref{Sec:ContactReductionField}) contains the field-theoretic analogue of the results of section (\ref{Sec:ContactReduction1}).\\

Finally, in a similar spirit to our treatment of the particle case, we conclude with a second complete example, in which we analyse a string-inspired, low-energy effective non-Abelian gauge theory \cite{dick1997implications}. The ideas developed over the course of this article constitute a foundational framework, within which to analyse singular (field) theories from a systematic perspective. It has recently been shown that the classical Einstein-Hilbert action possesses a redundant scaling degree of freedom, which is made manifest upon decomposing the spacetime metric into the product of a conformal factor and a symmetric rank-two tensor of fixed determinant \cite{sloan2025dynamical}. Further comments on the application of our work to General Relativity, together with a number of open questions, are presented in section (\ref{Sec:Conclusions}).
\section{Contact and Pre-contact Geometry}\label{Sec:ContactGeom}
In the context of analytical mechanics, symplectic geometry is, without doubt, the prevailing framework. Its even-dimensional manifolds and conserved phase space volumes make it the most appropriate formalism within which to construct a geometrical description of conservative mechanical systems. Contact geometry, by contrast receives far less attention within the literature, and so we dedicate this short section to a summary of the most relevant ideas. The interested reader is encouraged to consult \cite{de2019contact,de2020review,bravetti2019contact}.\\

In general, a contact manifold is an odd-dimensional manifold $C$, together with a maximally non-integrable distribution $\xi\subset TC$. Locally, this distribution may be described as the kernel of some $\eta\in\Omega^1(U\subset C)$, referred to as a contact form. If the quotient line bundle $TC/\xi \rightarrow C$ is trivial, as we shall assume, $\xi$ is said to be \textit{coorientable}, and $\eta$ may be extended to a global contact form on $C$. With the assumption of coorientability, we shall henceforth designate a contact manifold via the pair $(C,\eta)$. Every contact manifold admits a distinguished \textit{Reeb vector field} $\mathcal{R}\in\mathfrak{X}^{\infty}(C)$, defined via
\begin{equation}\label{Eq:Reeb}
    \iota_{\mathcal{R}}d\eta = 0 \quad\quad\quad\quad \iota_{\mathcal{R}}\eta =1
\end{equation}
As for symplectic manifolds, around each point $p\in C$, we may always find a local chart of Darboux coordinates $(x^1,\,\cdots,x^n,y_1,\,\cdots,y_n,z)$, in which $\eta$ takes the form $\eta = dz - y_i\,dx^i$. We shall refer to this as the \textit{canonical contact form}.\\

The extended cotangent bundle $T^*Q\times \mathbb{R}$ of the $n$-dimensional configuration space $Q$ is the canonical example of a contact manifold, upon which local coordinates are denoted $(q^i,p_i,S)$. Here, $S$ refers to the action, and as described in the introduction, the contact theories we shall study have the particular feature of being action-dependent. The canonical contact form on $T^*Q\times\mathbb{R}$ reads $\eta= dS-p_i\,dq^i$, and the Reeb vector field $\mathcal{R}$ follows from (\ref{Eq:Reeb}), and will often be taken to be $\mathcal{R}=\partial/\partial S$.\\

If $(TQ\times\mathbb{R},\eta_L,E_L)$ is a hyperregular Lagrangian system (that is to say, one for which the Legendre map $FL$ is a global diffeomorphism) we introduce the unique function $H:T^*Q\times\mathbb{R}\rightarrow\mathbb{R}$, such that $FL^*H=E_L$, in which $E_L$ is the Lagrangian energy function. The triple $(T^*Q\times\mathbb{R},\eta,H)$ is then said to constitute a contact Hamiltonian system, and there exists a bundle morphism
\begin{equation}\label{Eq:bHmorphism}
    \begin{split}
        \bar{\flat}\;:\; T(T^*Q\times\mathbb{R}) &\longrightarrow T^*(T^*Q\times \mathbb{R})\\
        v\;&\longmapsto \iota_v\,d\eta + (\iota_v\eta)\eta
    \end{split}
\end{equation}
The equations of motion are deduced seeking a vector field $X_{H}\in \mathfrak{X}^{\infty}(T^*Q\times\mathbb{R})$, which satisfies
\begin{equation}\label{Eq:GeometricalHamiltonEquations}
    \bar{\flat}(X_H) = dH - \left( \mathcal{R}(H)+H\right)\eta
\end{equation}
Suppose that we decompose the vector field $X_H$ as
\begin{equation}\label{Eq:XH}
    X_H = A^i\frac{\partial}{\partial q^i} + B_i\frac{\partial}{\partial p_i} + C\frac{\partial}{\partial S}
\end{equation}
then the coefficient functions $A^i$, $B_i$, and $C$ satisfy
\begin{equation}\label{Eq:LocalHamiltonEquations}
    A^i = \frac{\partial H}{\partial p_i} \quad\quad B_i = - \,\biggr(\frac{\partial H}{\partial q^i} + p_i \,\frac{\partial H}{\partial S}\biggr) \quad\quad C = p_i\, \frac{\partial H}{\partial p_i} - H
\end{equation}
\section{Contact Reduction of Singular Systems}\label{Sec:ContactReduction1}
Over the course of the following sections, we shall demonstrate how the contact reduction of Hamiltonian systems is to be modified when the system of interest must also be constrained, due to the presence of degeneracies. In order to facilitate this, we begin with a brief motivational discussion of scaling symmetries and contact reduction in a more general setting.
\subsection{Scaling Symmetries of Regular Hamiltonian Systems}\label{Subsec:ScalingSymmetries}
In general, mechanical systems possess a multitude of degrees of freedom, the dynamics of which encode how the system (or parts of) evolve over time. We consider an `observable' to be any quantity that is accessible via empirical methods to somebody within the system; it is frequently the case that these observables form a dynamical algebra (with operations of addition and multiplication inherited from those defined on phase space). Within this algebra, there may also be a number of unobservable quantities, whose presence is required to make the algebra close. In the introduction, we described how scaling symmetries indicate the existence of degrees of freedom that play no role in the closure of the dynamical algebra of observables. Indeed such parameters often arise when one constructs a mathematical framework which introduces superfluous structure, such as external rods and clocks, whose sole purpose is to expedite practical computations.\\

We propose that such an approach to theorising is misguided, and justify our stance from various physical and philosophical standpoints. In addition to Leibniz's principle of the Identity of Indiscernibles, to which we made extensive reference in the introduction, we also appeal to the work of Ismael and van Fraassen \cite{ismael2021symmetry,ismael2003symmetry}. Here, it is argued that when attempting to construct a theory of the natural world, the resulting model should be minimally sufficient to describe the phenomena of interest, without adding any superfluous elements - even if such elements provide a means to make practical calculations more streamlined. From a physical perspective, there are also several reasons one should advocate for a minimalistic approach, the majority of which are exemplified by the field of cosmology. When we study a system in the lab, we may always justify the introduction of external rods and clocks, claiming that they are external to the system, but accessible to us as observers. Such reasoning fails when the system of interest happens to be the universe as a whole - of what kind of apparatus may we conceive, which resides outside of the universe? A further motivating factor is that classical General Relativity ceases to provide reliable predictions at points where the spacetime geometry becomes singular. In many instances, the resulting pathologies are superficial - an artifice of the additional structure, rather than a breakdown of the underlying physical theory.\\

A particularly simple example of this is the case of a (flat) FLRW cosmology with minimally-coupled scalar matter. Typically, the scale factor plays a central role in the description of these models, and yet it is known that the value of $a(t)$ itself carries no physical meaning. Inferences about the underlying spacetime geometry are made by observing how the matter evolves over time. The only way in which the scale factor encodes physically observable information is through \textit{ratios} of its values at different times. In this way, $a(t)$ is a redundant degree of freedom within the cosmological description. Further, the association of the scale factor with an overall notion of size becomes highly problematic as one approaches the initial singularity. Indeed, it has been shown that by working with a contact-reduced system, which makes no reference to $a(t)$, solutions may be continued predictively beyond this point \cite{sloan2019scalar}.\\

After this somewhat lengthy motivational introduction, we turn to a more mathematical discussion of scaling symmetries and contact reduction. Examples and further comments may be found in \cite{bravetti2023scaling,sloan2021scale}, for example. Given a symplectic Hamiltonian system, $(M,\omega,H)$, we declare a vector field $D\in\mathfrak{X}^{\infty}(M)$ to constitute a scaling symmetry of degree $\Lambda\in\mathbb{R}$ if the following two conditions hold
\begin{equation}\label{Eq:DSConditions}
    \mathfrak{L}_D\hspace{0.2mm}\omega=\omega\quad\quad\quad\quad \mathfrak{L}_DH=\Lambda H
\end{equation}
in which $\mathfrak{L}$ denotes the Lie derivative. As a consequence, we see that if $X_H\in\mathfrak{X}^{\infty}(M)$ is the unique Hamiltonian vector field corresponding to $H$ via $\iota_{X_H}\omega=dH$, then
\begin{equation}
    \iota_{\scriptscriptstyle[D,X_H]}\hspace{0.3mm}\omega = [\mathfrak{L}_D,\iota_{X_H}]\hspace{0.4mm}\omega = (\Lambda-1) \,dH
\end{equation}
Since $\omega$ is, by assumption, non-degenerate, it follows that $[D,X_H]=(\Lambda-1)X_H$, and the generator $D$ acts only to rescale the Hamiltonian flow. A non-trivial corollary of this is that the equations of motion of the invariants of $D$, i.e those quantities $\phi$ such that $\mathfrak{L}_D\phi=D[\phi]=0$, are unaffected by the action of $D$. Further, these invariants are found to form an algebra \cite{sloan2018dynamical}.\\

If the vector field $D$ is such that its flow acts freely and properly on $M$, then the quotient space $M/\hspace{-1mm}\sim$ formed by identifying points connected by $D$-orbits has the structure of a smooth manifold of dimension $\textrm{dim}\,M-1$; moreover, the map $\pi:M\rightarrow M/\hspace{-1mm}\sim$, sending each point of $M$ to its equivalence class, is a submersion \cite{bravetti2019contact}. In addition to being a smooth manifold, the quotient space also inherits a contact structure $\xi:= \pi_*\,\textrm{ker}\left(\iota_D\hspace{0.2mm}\omega\right)$. We define a scaling function to be any $\rho:M\rightarrow\mathbb{R}_+$, such that $\mathfrak{L}_D\rho=\rho\,$; the existence of a global scaling function allows the contact distribution $\xi$ to be expressed as the kernel of a well-defined 1-form $\eta$ on $M/\hspace{-1mm}\sim\,$
\begin{equation}\label{Eq:ContactForm}
    \pi^*\eta:=\frac{\iota_D\hspace{0.2mm}\omega}{\rho}
\end{equation}
Assuming the existence of a global scaling function, we adopt the notation $(C,\eta)$ to refer to the contact manifold $M/\hspace{-1mm}\sim$. Orbits of the symplectic Hamiltonian system project to curves on $C$; more precisely, the Hamiltonian vector field $X_H\in\mathfrak{X}^{\infty}(M)$ generates a line field $\textrm{span}\left(X_H\right)$ on $M$, which is projected to a line field $\pi_*\,\textrm{span}\left(X_H\right)$ on $C$. Additionally, there exists a contact Hamiltonian function $H^c:C\rightarrow\mathbb{R}$, which, on the level-set $\pi(H=0)$, is calculated according to
\begin{equation}\label{Eq:ContactHamiltonian}
    \pi^*H^c:=\frac{H}{\rho^{\Lambda}}
\end{equation}
Elsewhere, i.e on $C\hspace{0.6mm}\backslash \pi(H=0)$, the appropriate function is $|H^c|^{1/\Lambda}$. If $X_{H^c}$ denotes the Hamiltonian vector field of $H^c$, and $\mathcal{R}$ is the Reeb field of $\eta$, then the vector field 
\begin{equation}\label{Eq:XField}
    X:= X_{H^c} + (\Lambda-1)H^c\,\mathcal{R}
\end{equation}
generates $\pi_*\,\textrm{span}(X_H)$. A contact vector field is one whose flow preserves the contact distribution. Unless our scaling symmetry is of degree one (which is often the case) or we restrict ourselves to the zero-set of $H^c$, the vector field $X$ does not generally preserve $\xi=\textrm{ker}\,\eta$. Further, the presence of the scaling function $\rho$ introduces a temporal reparameterisation on $C$. For a scaling symmetry of degree $\Lambda$, trajectories on the symplectic phase space, parameterised by $t$, are projected to curves governed by $\tau$, with $d\tau=\rho^{\Lambda-1}dt$. Accordingly, if $(q^i(\tau),p_i(\tau),S(\tau))$ describes an integral curve of (\ref{Eq:XField}), the equations of motion read
\begin{equation}\label{Eq:ReparameterisedHamiltonEq}
    \frac{dq^i}{d\tau} = \frac{\partial H^c}{\partial p_i} \quad\quad \frac{dp_i}{d\tau} = -\left(\frac{\partial H^c}{\partial q^i} + p_i\frac{\partial H^c}{\partial S}\right) \quad\quad \frac{dS}{d\tau} = p_i\frac{\partial H^c}{\partial p_i} - \Lambda H^c
\end{equation}
This concludes our introductory treatment of contact reduction, and we now apply these ideas to constrained Hamiltonian systems. Note that there are two ways to proceed: given a Hamiltonian $H_0$, corresponding to a singular Lagrangian, we may deduce the final constraint manifold, and then make the contact reduction, or identify a scaling degree of freedom within the unconstrained Hamiltonian system, excise this from our ontology, and then use the pre-contact constraint algorithm.
\subsection{Singular Systems: Reduction $\&$ Restriction}\label{Subsec:Reduction+Restriction}
A Lagrangian system $(TQ,\omega_L,E_L)$ is said to be \textbf{almost-regular} if
\begin{itemize}
    \item[$\star$] $M_0:= FL(TQ)$ is a closed submanifold of $T^*Q$
    \item[$\star$] $FL$ is a submersion onto its image
    \item[$\star$] For every $p\in M_0$, the fibres $FL^{-1}(p)$ are connected submanifolds of $TQ$
\end{itemize}
Given such a system, we begin by deducing the canonical Hamiltonian $H_0:M_0\rightarrow \mathbb{R}$, defined on the primary constraint manifold $\jmath_0:M_0\hookrightarrow T^*Q$. This is then extended to a function $H:T^*Q\rightarrow\mathbb{R}$ on the full phase space, such that $H|_{M_0}=H_0$. We remark, for future reference, that a quantity is said to be \textbf{weakly-vanishing} if it is zero when restricted to the constraint surface. In local Darboux coordinates, the canonical symplectic form on $T^*Q$ reads $\omega=dq^i\wedge dp_i$. If $M_0$ is described as the zero-set of the primary constraint functions $\phi^{\alpha}$ (these arise as a direct consequence of the non-invertibility of $FL$, rather than as a result of using the equations of motion) we seek a vector field $D\in\mathfrak{X}^{\infty}(T^*Q)$ which satisfies
\begin{equation}
    \mathfrak{L}_DH=\Lambda H \quad\quad\quad \mathfrak{L}_D\hspace{0.2mm}\omega=\omega \quad\quad\quad \mathfrak{L}_D\phi^{\alpha}= C^{\alpha}_{\,\beta} \phi^{\beta}
\end{equation}
for a set of functions $C^{\alpha}_{\,\beta}\in C^{\infty}(T^*Q)$, with $\textrm{det}\,C^{\alpha}_{\;\beta}\neq 0$. The first two conditions identify $D$ as a scaling symmetry, while the third ensures that the flow of $D$ maps $M_0$ onto itself. This is essential, as $M_0$ describes the maximal subset of points at which solutions to the equations of motion could exist; the flow of $D$ must not, therefore, map points of $M_0$, which are (potentially) dynamically admissible, to points of $T^*Q\hspace{0.8mm}\backslash M_0$, which are not.\\

In principle, we should now verify that the flow of $D$ acts freely and properly on $T^*Q$; supposing that this \textit{is} the case, the space $C:=T^*Q/\hspace{-1mm}\sim$ is a contact manifold with submersion $\beta:T^*Q\rightarrow C$, and contact distribution $\xi:= \beta_*\,\textrm{ker}\left(\iota_D\hspace{0.2mm}\omega\right)$. The contact form $\eta$ and Hamiltonian $H^c$ are defined precisely as in (\ref{Eq:ContactForm}) and (\ref{Eq:ContactHamiltonian}) respectively, with the result that $(C,\eta,H^c)$ constitutes a contact Hamiltonian system. We highlight that this construction does not directly produce a pre-contact manifold, as we have symmetry-reduced the full phase space, into which the pre-symplectic system is embedded. The map $\beta: T^*Q\rightarrow C$ allows us to construct functions $\gamma^{\alpha}\in C^{\infty}(C)$, whose zero-set defines a submanifold $\kappa_0:C_0\hookrightarrow C$, which \textit{does} inherit a pre-contact structure; in particular, we have 
\begin{equation}\label{Eq:C0}
    C_0:=\{y\in C\;|\; \gamma^{\alpha}(y)=0\,\}\quad\quad\textrm{with}
    \quad \beta^*\gamma^{\alpha}:=\frac{\phi^{\alpha}}{\rho^{\Lambda}}
\end{equation}
The contact Hamiltonian $H^c$ is restricted to a function $H^c_0:=H^c|_{C_0}$, and similarly, $\eta_0:=\eta|_{C_0}$ defines a pre-contact form on $C_0$, with the result that the triple $(C_0,\eta_0,H^c_0)$ is a well-defined pre-contact manifold, to which the geometric constraint algorithm may freely be applied. While constructed in detail in \cite{bell2026constrained}, we briefly recall that the pre-contact algorithm seeks a vector field solution to the contact Hamiltonian equation of motion. At each stage, we have a constraint submanifold $\kappa_{i}:C_{i}\hookrightarrow C_{i-1}$, which is some restricted subset of $C_0$. Each stage of the algorithm requires that the vector field solution be tangent to the constraint surface, else the dynamics would tend to evolve off this surface, and the system would cease to respect the constraints. Recursively imposing this tangency condition is what produces the series of embedded submanifolds $\kappa_{i+1}:C_{i+1}\hookrightarrow C_{i}$. Supposing that the algorithm stabilises on the final constraint surface $C_f$, we eliminate any gauge degrees of freedom, obtaining a physical state space $C_{\scriptscriptstyle\mathcal{P}}$, with $\kappa_{\scriptscriptstyle\mathcal{P}}:C_{\scriptscriptstyle\mathcal{P}}\hookrightarrow C_f$, which inherits a contact form $\eta_{\scriptscriptstyle\mathcal{P}}$, and Hamiltonian $H^c_{\scriptscriptstyle\mathcal{P}}$.
\subsection{Singular Systems: Restriction $\&$ Reduction}\label{Subsec:Restriction+Reduction}
In order to carry out the process in which we seek a scaling symmetry within the constrained theory, we must apply the pre-symplectic constraint algorithm to the system $(M_0,\omega_0,H_0)$, which produces a series of submanifolds $\jmath_{i+1}:M_{i+1}\hookrightarrow M_i$, and terminates on the final constraint surface $M_f$. We must then either make a choice of gauge-fixing, or formally quotient out by the action of the gauge transformations. Since we are concerned with the dynamics of observable, gauge-invariant quantities, this choice is somewhat inconsequential, and we denote the physical phase space $\mathcal{P}$, with $\jmath_{\scriptscriptstyle\mathcal{P}}:\mathcal{P}\hookrightarrow M_f$. This space is \textit{symplectic}, with non-degenerate 2-form $\omega_{\scriptscriptstyle\mathcal{P}}$ and Hamiltonian $H_{\scriptscriptstyle\mathcal{P}}$. As such, we now seek a vector field $Z\in\mathfrak{X}^{\infty}(\mathcal{P})$, such that
\begin{equation}\label{Eq:Deg}
    \mathfrak{L}_Z\omega_{\scriptscriptstyle\mathcal{P}}=\omega_{\scriptscriptstyle\mathcal{P}}\quad\quad\quad \mathfrak{L}_ZH_{\scriptscriptstyle\mathcal{P}} = \Lambda H_{\scriptscriptstyle\mathcal{P}}
\end{equation}
Note that, by construction, the constraint algorithm produces a surface to which all dynamics remain tangent; thus, we need impose no conditions on the preservation of the constraints. The dynamical evolution of some $f\in C^{\infty}(\mathcal{P})$ is determined via the bracket induced by $\omega_{\scriptscriptstyle\mathcal{P}}$, that is
\begin{equation*}
    \dot{f} = \{f,H_{\scriptscriptstyle\mathcal{P}}\}_{\scriptscriptstyle\mathcal{P}} := \omega_{\scriptscriptstyle\mathcal{P}}(X_f,X_{H_{\scriptscriptstyle\mathcal{P}}})
\end{equation*}
in which $X_f$ refers to the Hamiltonian vector field associated to $f$ via $\iota_{X_f}\omega_{\scriptscriptstyle\mathcal{P}}=df$, and similarly for $X_{H_{\scriptscriptstyle\mathcal{P}}}$. Assuming the flow of $Z$ acts freely and properly on $\mathcal{P}$, we have the quotient space $\mathcal{P}/\hspace{-1mm}\sim$, with submersion $\sigma:\mathcal{P}\rightarrow\mathcal{P}/\hspace{-1mm}\sim\,$; if $\rho:\mathcal{P}\rightarrow\mathbb{R}_+$ is a global scaling function, the contact form and Hamiltonian read
\begin{equation}\label{Eq:EtaandHcAgain}
    \sigma^*\eta_{\scriptscriptstyle\mathcal{P}} = \frac{\iota_Z\omega_{\scriptscriptstyle\mathcal{P}}}{\rho}\quad\quad\quad\sigma^*H_{\scriptscriptstyle\mathcal{P}}^c = \frac{H_{\scriptscriptstyle\mathcal{P}}}{\rho^{\Lambda}}
\end{equation}
Supposing that $(q^i,\Pi_i,S)$ constitutes a set of local Darboux coordinates on $\mathcal{P}/\hspace{-1mm}\sim$, so that $\eta_{\scriptscriptstyle\mathcal{P}}=dS-\Pi_i dq^i$, the symmetry-reduced equations of motion are simply the contact Hamiltonian equations (\ref{Eq:ReparameterisedHamiltonEq}).\\

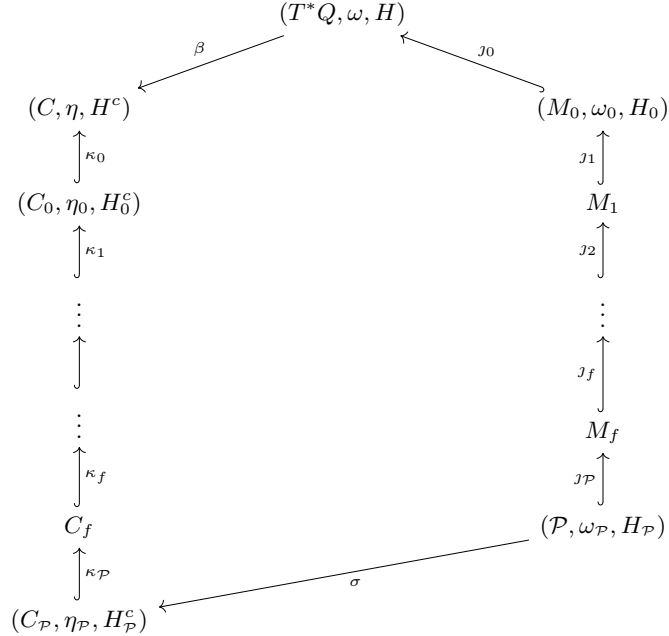
\begin{figure}
\begin{tikzcd}[column sep=4.5em, row sep=2.2em]
& (T^*Q,\omega,H) \arrow[dl, "\beta"']  & \\
(C,\eta,H^c)
    & 
    &
    (M_0,\omega_0,H_0) \arrow[hookrightarrow,ul, "\jmath_0"'] \\ 
(C_0,\eta_0,H^c_0) \arrow[hookrightarrow,u,"\kappa_0"'] 
    & 
    &
    M_1 \arrow[hookrightarrow,u, "\jmath_1"] \\ 
\vdots \arrow[hookrightarrow,u,"\kappa_1"'] 
    & 
    &
    \vdots \arrow[hookrightarrow,u,"\jmath_2"] \\
\vdots \arrow[hookrightarrow, u] 
    & 
    &
    M_f \arrow[hookrightarrow,u,"\jmath_f"] \\
C_f \arrow[hookrightarrow,u,"\kappa_f"']
    &
    &
    (\mathcal{P},\omega_{\scriptscriptstyle\mathcal{P}},H_{\scriptscriptstyle\mathcal{P}}) \arrow[hookrightarrow,u,"\jmath_{\scriptscriptstyle\mathcal{P}}"] \arrow[dll, "\sigma"] \\
(C_{\scriptscriptstyle\mathcal{P}},\eta_{\scriptscriptstyle\mathcal{P}},H^c_{\scriptscriptstyle\mathcal{P}}) \arrow[hookrightarrow,u,"\kappa_{\scriptscriptstyle\mathcal{P}}"']
\end{tikzcd}
\caption{Commutative diagram showing how the two constraint algorithms run in parallel. Starting from the cotangent bundle, the right-hand path shows implementation of the pre-symplectic algorithm, to deduce the final space $\mathcal{P}$. This manifold is symplectic, and allows a contact reduction to be carried out. Alternatively, the scaling variable may be excised from the full symplectic phase space, defining the contact manifold $C$. Upon projecting the primary constraint functions that define $M_0$, $C$ is restricted to the \textit{pre-contact} space $C_0$, and the left-hand path depicts the implementation of the pre-contact algorithm. Both paths lead to the same final reduced space $C_{\scriptscriptstyle\mathcal{P}}$.}
\label{Fig:figure1}
\end{figure}

From the constructions given above, it should be clear that the pre-symplectic and pre-contact constraint algorithms run in parallel. At the $i^{\textrm{th}}$ stage of the former, the submanifold $M_i$, with inclusion $\jmath_i:M_i\hookrightarrow M_{i-1}$, may be described as the zero-set of the functions $\phi^{\alpha}_{(i)}$, whose projection defines a second set of functions $\gamma^{\alpha}_{(i)}$
\begin{equation}\label{Eq:GammaConstraints}
    \beta^*\gamma^{\alpha}_{(i)}:=\frac{\phi^{\alpha}_{(i)}}{\rho^{\Lambda}}
\end{equation}
Up to multiplicative factors, these $\gamma^{\alpha}_{(i)}$ coincide with the constraint functions obtained in $i^{\textrm{th}}$ iteration of the pre-contact algorithm. These ideas are summarised in figure \ref{Fig:figure1}.\\
\section{An Example}\label{Sec:ParticleEg1}
We now apply the formalism developed throughout to a simple example; for completeness, we obtain the reduced space dynamics via both paths of figure \ref{Fig:figure1}, providing an explicit illustration of the commutativity of the reduction and constraint processes.
\subsection{Restriction $\&$ Reduction}
For this particular example, we take $Q=\mathbb{R}_+^4$, with local coordinates $(x,y,u,\phi)$, and Lagrangian function
\begin{equation}\label{Eq:Eg1Lagrangian}
    L = 2\dot{u}^2+ \frac{1}{2}u^2\biggr[\dot{\phi}^2 + (\dot{x}+\dot{y}-\dot{\phi})^2-(x^2+y^2-2\phi^2)\biggr]
\end{equation}
Our motivation for choosing this particular Lagrangian is purely pedagogical; we do not consider this to represent any realistic physical system. Clearly, the Lagrangian is singular (in the sense of the discussion of the introduction), as the Hessian matrix with respect to the velocities $(\dot{x},\dot{y},\dot{u},\dot{\phi})$ is
\begin{equation*}
    W = u^2\begin{pmatrix}
        1 & 1 & 0 & -1\\
        1 & 1 & 0 & -1\\
        0 & 0 & 4u^{-2} & 0\\
        -1 & -1 & 0 & 2
    \end{pmatrix}
\end{equation*}
which is of constant rank 3, since $u\neq0$. The non-invertibility of this matrix indicates that, as expected, we have a singular system. Making the change of variable $u\mapsto e^{\rho/2}$, the Lagrangian reads
\begin{equation}\label{Eq:Eg1RhoLagrangian}
    L = e^{\rho}\biggr[\frac{1}{2}\left(\dot{\rho}^2+\dot{\phi}^2\right) + \frac{1}{2}\left(\dot{x}+\dot{y}-\dot{\phi}\right)^2-\frac{1}{2}\left(x^2+y^2\right) +\phi^2\biggr]
\end{equation}
This is the most convenient parameterisation of $L$, and the one with which we shall henceforth work. The momenta conjugate to $(\dot{x},\dot{y},\dot{u},\dot{\phi})$ are given by
\begin{align*}
    p_{\rho} & = e^{\rho}\dot{\rho} & p_x & = e^{\rho}\left(\dot{x}+\dot{y}-\dot{\phi}\right)\\
    p_{\phi} &= e^{\rho}\left(2\dot{\phi} - (\dot{x}+\dot{y})\right) & p_y & = e^{\rho}\left(\dot{x}+\dot{y}-\dot{\phi}\right)
\end{align*}
From this, we see that there is a single primary constraint, defining the submanifold $\jmath_0:M_0\hookrightarrow T^*Q$
\begin{equation}\label{Eq:Eg1M0}
    M_0:=\{p\in T^*Q\;|\; \psi^1(p):=p_x-p_y = 0\,\}
\end{equation}
The canonical Hamiltonian $H_0:M_0\rightarrow \mathbb{R}$ is readily calculated to be 
\begin{equation}\label{Eq:Eg1H0}
    H_0=\frac{e^{-\rho}}{2}\left[p_{\rho}^2 + \left(p_{x}+p_{\phi}\right)^2+p_x^2\right] + e^{\rho}\left[\frac{1}{2}(x^2+y^2) -\phi^2\right]
\end{equation}
Pulling back the canonical symplectic form $\omega$ on $T^*Q$ to $M_0$, we have
\begin{equation}\label{Eq:Eg1omega0}
    \omega_0 = d\rho\wedge dp_{\rho} + d\phi\wedge dp_{\phi} + (dx + dy)\wedge dp_x
\end{equation}
Applying the steps of the pre-symplectic constraint procedure presented in \cite{bell2026constrained}, we find that the algorithm stabilises after a single iteration, yielding a final submanifold
\begin{equation}\label{Eq:Eg1M1}
    M_1 = M_f = \{p\in T^*Q\;|\; \psi^1(p) = p_x-p_y = 0\;;\; \psi^2(p)= e^{\rho}\left(x-y\right) = 0\,\}
\end{equation}
Introducing the matrix of Poisson brackets $J^{\alpha\beta}:=\{\psi^{\alpha},\psi^{\beta}\}$, we deduce that $\psi^1$ and $\psi^2$ are both second-class (non-weakly-vanishing Poisson bracket with some or all other constraints), as $J$ is invertible; indeed
\begin{equation*}
    J = e^{\rho} \begin{pmatrix}
        0 & -2\\
        2 & 0
    \end{pmatrix}\hspace{0.7cm}\implies \hspace{0.7cm} J^{-1} = \frac{e^{-\rho}}{2}\begin{pmatrix}
        0 & 1\\
        -1 & 0
    \end{pmatrix}
\end{equation*}
With this, there are no gauge degrees of freedom, and so we write $M_f=\mathcal{P}$. Since we have only second-class constraints, these may be imposed as strong equalities. In order to be able to do this \textit{before} computing the equations of motion, we must introduce a modification of the Poisson bracket, known as the Dirac bracket. In the current context, the Dirac bracket of two functions $f,g\in C^{\infty}(T^*Q)$ reads
\begin{equation}\label{Eq:Eg1Dirac}
    \{f,g\}_D = \{f,g\} - \frac{e^{-\rho}}{2}\{f,\psi^1\}\{\psi^2,g\} + \frac{e^{-\rho}}{2} \{f,\psi^2\}\{\psi^1,g\}
\end{equation}
The equations of motion for our phase space variables then follow from $\dot{f}= \{f,H_0\}_D|_{\scriptscriptstyle \mathcal{P}}$, and we find
\begin{align}\label{Eq:Eg1SympEoM}
    \dot{x} & = \frac{e^{-\rho}}{2}\left(2p_x+p_{\phi}\right) = \dot{y} & \dot{p}_x &= -e^{\rho} x = \dot{p}_y\notag\\
    \dot{\phi} &= e^{-\rho}\left(p_x+p_{\phi}\right) & \dot{p}_{\phi} & = 2e^{\rho}\phi\\
    \dot{\rho} &=e^{-\rho}p_{\rho} & \dot{p}_{\rho} &= \frac{e^{-\rho}}{2}\left(p_{\rho}^2+\left(p_x+p_{\phi}\right)^2+p_x^2\right) - e^{\rho}\left(x^2-\phi^2\right)\notag
\end{align}
These are the dynamical equations that must be reproduced by the symmetry-reduced model. Now, note that the symplectic form on $\mathcal{P}$ is
\begin{equation*}
    \omega_{\scriptscriptstyle\mathcal{P}}:= \omega_0|_{\scriptscriptstyle \mathcal{P}} = d\rho \wedge dp_{\rho} + d\phi\wedge dp_{\phi}+ 2\,dx\wedge dp_x
\end{equation*}
We also have the following restricted Hamiltonian
\begin{equation*}
    H_{\scriptscriptstyle\mathcal{P}}:= H_0|_{\scriptscriptstyle \mathcal{P}} = \frac{e^{-\rho}}{2}\left[p_{\rho}^2 + \left(p_{x}+p_{\phi}\right)^2+p_x^2\right] + e^{\rho}\left[x^2 -\phi^2\right]
\end{equation*}
In accordance with the results of section (\ref{Subsec:Restriction+Reduction}), we now seek a vector field $Z\in\mathfrak{X}^{\infty}(\mathcal{P})$ that satisfies $\mathfrak{L}_Z\hspace{0.3mm}\omega_{\scriptscriptstyle\mathcal{P}}=\omega_{\scriptscriptstyle\mathcal{P}}$, and $\mathfrak{L}_Z\hspace{0.2mm}H_{\scriptscriptstyle\mathcal{P}}=\Lambda H_{\scriptscriptstyle\mathcal{P}}$; a series of short calculations confirms that
\begin{equation}
    Z = \frac{\partial}{\partial \rho} + p_{\rho}\frac{\partial}{\partial p_{\rho}} + p_{\phi}\frac{\partial}{\partial p_{\phi}} + p_{x}\frac{\partial}{\partial p_{x}}
\end{equation}
satisfies both requirements, and is a scaling symmetry of degree one, i.e. it satisfies the conditions of (\ref{Eq:Deg}) with $\Lambda=1$. An obvious choice of scaling function is $e^{\rho}$; thus, we introduce the map $\sigma: \mathcal{P}\rightarrow \mathcal{P}/\hspace{-1mm}\sim$, and construct the 1-form and contact Hamiltonian as
\begin{equation*}
    \sigma^*\eta_{\scriptscriptstyle\mathcal{P}} = \frac{\iota_Z\hspace{0.2mm}\omega_{\scriptscriptstyle\mathcal{P}}}{e^{\rho}} \quad\quad \sigma^*H_{\scriptscriptstyle\mathcal{P}}^c = \frac{H_{\scriptscriptstyle\mathcal{P}}}{e^{\rho}}
\end{equation*}
We take coordinates on $\mathcal{P}/\hspace{-1mm}\sim$ to be $(x,\Pi_x,\phi,\Pi_{\phi},S)$, where
\begin{equation}\label{Eq:Eg1ReducedCoords1}
    \sigma^* \Pi_x = \frac{p_x}{e^{\rho}} \quad\quad \sigma^* \Pi_{\phi} = \frac{p_{\phi}}{e^{\rho}} \quad\quad \sigma^* S = \frac{p_{\rho}}{e^{\rho}}
\end{equation}
It then follows that $\eta_{\scriptscriptstyle\mathcal{P}}$ and $H_{\scriptscriptstyle\mathcal{P}}^c$ have the following local coordinate expressions
\begin{equation}\label{Eq:Eg1ContactEtaandH1}
    \eta_{\scriptscriptstyle\mathcal{P}} = dS-\Pi_{\phi} d\phi - 2\hspace{0.2mm}\Pi_x dx\quad\quad\quad H_{\scriptscriptstyle\mathcal{P}}^c = \frac{1}{2}\biggr[S^2 +\left(\Pi_x+\Pi_{\phi}\right)^2+\Pi_x^2\biggr] +x^2-\phi^2
\end{equation}
The contact Hamiltonian equations of motion are deduced in the standard fashion; however, it should be noted that the multiplicative factor of two in our expression for $\eta_{\scriptscriptstyle\mathcal{P}}$ shows that $(x,2\Pi_x,\phi,\Pi_{\phi},S)$ provides the correct set of local Darboux coordinates on $\mathcal{P}/\hspace{-1mm}\sim$. Consequently, we have
\begin{align}\label{Eq:Eg1ContactEquations1}
    \dot{x} &= \frac{1}{2}\left(2\Pi_x+\Pi_{\phi}\right) & \dot{\Pi}_x &= -\left(2x + \Pi_x S\right)\notag \\
    \dot{\phi} &=\Pi_x+\Pi_{\phi} & \dot{\Pi}_{\phi} &= 2\phi-\Pi_{\phi}S\\
    \dot{S} &= \frac{1}{2}\biggr[-S^2 + \left(\Pi_x+\Pi_{\phi}\right)^2 \biggr] - x^2+\phi^2 &\notag
\end{align}
It is clear that the equations of motion for the coordinates $x$ and $\phi$ coincide with those given in (\ref{Eq:Eg1SympEoM}); after some work, it can be shown that the additional action-dependent terms in the momentum equations correctly reproduce the original dynamics.
\subsection{Reduction + Restriction}
In order to obtain the contact Hamiltonian equations (\ref{Eq:Eg1ContactEquations1}) via the alternative method presented in section (\ref{Subsec:Reduction+Restriction}), we return to the canonical Hamiltonian $H_0$ on the primary constraint manifold; in accordance with our general procedure, we extend $H_0$ to a function $H$ on the full phase space $T^*Q$
\begin{equation}\label{Eq:ExtendedH}
    H=\frac{e^{-\rho}}{2}\left[p_{\rho}^2 + \left(p_{x}+p_{\phi}\right)^2+p_x^2\right] + e^{\rho}\left[\frac{1}{2}(x^2+y^2) -\phi^2\right]
\end{equation}
The vector field
\begin{equation}
    D = \frac{\partial}{\partial \rho} + p_{\rho}\frac{\partial}{\partial p_{\rho}} + p_{\phi}\frac{\partial}{\partial p_{\phi}} + p_{x}\frac{\partial}{\partial p_{x}} + p_{y}\frac{\partial}{\partial p_{y}}
\end{equation}
is a scaling symmetry of degree one, and, crucially, preserves the primary constraint $\psi^1$, since $\mathfrak{L}_D\psi^1=\psi^1$. Thus, we know that the quotient space $C:=T^*Q/\hspace{-1mm}\sim$ is a contact manifold, and we introduce the surjective map $\beta:T^*Q\rightarrow C$, and take coordinates on $C$ to be $(x,\pi_x,y,\pi_y,\phi,\pi_{\phi},S)$, where
\begin{equation}
    \beta^*\pi_x=\frac{p_x}{e^{\rho}} \quad\quad\quad \beta^*\pi_y=\frac{p_y}{e^{\rho}} \quad\quad\quad \beta^*\pi_{\phi}=\frac{p_{\phi}}{e^{\rho}} \quad\quad\quad \beta^*S = \frac{p_{\rho}}{e^{\rho}}
\end{equation}
so that the contact form and Hamiltonian on $C$ become
\begin{equation*}
    \eta = dS - \pi_x dx - \pi_y dy -\pi_{\phi} d\phi \quad\quad\quad H^c = \frac{1}{2}\biggr[S^2+\left(\pi_x+\pi_{\phi}\right)^2 +\pi_x^2\biggr] + \frac{1}{2}\left(x^2+y^2\right)-\phi^2
\end{equation*}
We emphasise that $(C,\eta,H^c)$ is a \textit{contact} Hamiltonian system, and that the symmetry-reduced analogue of the primary constraint manifold $M_0$ is found by projecting the function $\psi^1$, to obtain
\begin{equation*}
    \beta^*\gamma^1 = \frac{\psi^1}{e^{\rho}} \quad \implies \quad \gamma^1 = \pi_x + \pi_y
\end{equation*}
Consequently, the primary constraint submanifold $C_0$ is precisely the zero-set of $\gamma^1$, and the pre-contact system $(C_0,\eta_0,H^c_0)$ is obtained, restricting $\eta$ and $H^c$ to this space
\begin{equation*}
    \eta_0:=\eta|_{C_0} = dS - \pi_x\left(dx+dy\right) -\pi_{\phi}d\phi \quad\quad\quad H^c_0:=H^c|_{C_0} = \frac{1}{2}\biggr[S^2+\left(\pi_x+\pi_{\phi}\right)^2 +\pi_x^2\biggr] + \frac{1}{2}\left(x^2+y^2\right)-\phi^2
\end{equation*}
A single iteration of the pre-contact constraint algorithm gives rise to a secondary constraint
\begin{equation*}
    C_1=\{ p\in C_0\;|\; \gamma^2(p):= x-y = 0\,\}
\end{equation*}
As expected, comparing this space to $M_1$ found in (\ref{Eq:Eg1M1}), we see that the constraint $\gamma^2$ satisfies
\begin{equation*}
    \beta^*\gamma^2 = \frac{\psi^2}{e^{\rho}}
\end{equation*}
Further iteration produces no additional constraints, and so the algorithm stabilises on $C_1$. In order to compute the equations of motion, we require the matrix of Jacobi brackets \cite{bell2026constrained} $K^{\alpha\beta}:=\{\gamma^{\alpha},\gamma^{\beta}\}_J$
\begin{equation*}
    K = \begin{pmatrix}
        0 & -2\\
        2 & 0
    \end{pmatrix} \quad\quad\implies\quad\quad K^{-1} = \frac{1}{2}\begin{pmatrix}
        0 & 1\\
        -1 & 0
    \end{pmatrix}
\end{equation*}
The invertibility of this matrix confirms that, as expected, $\gamma^1$ and $\gamma^2$ are both second-class constraints; thus, in order to calculate the equations of motion, we impose such conditions as strong equalities, and introduce the Dirac-Jacobi bracket. The Dirac-Jacobi bracket is the contact-analogue of the Dirac bracket, which we introduced above for symplectic systems. Here, this Dirac-Jacobi bracket reads
\begin{equation}\label{Eq:Eg1DJBracket}
    \{f,g\}_{DJ} = \{f,g\}_J - \frac{1}{2}\{f,\gamma^1\}_J\,\{\gamma^2,g\}_J + \frac{1}{2}\{f,\gamma^2\}_J\,\{\gamma^1,g\}_J
\end{equation}
The evolution of a function $f$ is then computed from $\{f,H^c_0\}_{DJ}\,|_{\scriptscriptstyle C_1}$, and we find that
\begin{align}\label{Eq:Eg1ContactEquations2}
    \dot{x} &= \frac{1}{2}\left(2\pi_x+\pi_{\phi}\right) & \dot{\pi}_x &= -\left(2x + \pi_x S\right)\notag \\
    \dot{\phi} &=\pi_x+\pi_{\phi} & \dot{\pi}_{\phi} &= 2\phi-\pi_{\phi}S\\
    \dot{S} &= \frac{1}{2}\biggr[-S^2 + \left(\pi_x+\pi_{\phi}\right)^2 \biggr] - x^2+\phi^2 &\notag
\end{align}
which coincides precisely with the corresponding set of expressions (\ref{Eq:Eg1ContactEquations1}), deduced by first restricting the pre-symplectic system, and then making a symmetry reduction. In this example, we have provided an explicit illustration of how a constrained Hamiltonian system may be reduced to a simpler, dynamically-equivalent theory, by identifying and excising a superfluous scaling degree of freedom. Whilst relatively simple in nature, the system considered has allowed us to present a fully-worked example of the commutativity of contact reduction and phase space restriction, and thus concludes our treatment of particle dynamics.
\section{Multisymplectic Field Theory}\label{Sec:MultisymplecticFieldTheory}
The use of multisymplectic geometry to describe classical field theories is an area of active research interest, as the fibred manifolds it employs provide an arena in which a manifestly covariant formalism may be developed, in a finite-dimensional setting \cite{cantrijn1999geometry,forger2013multisymplectic}. Despite this, the multisymplectic Hamiltonian formalism of singular field theories is still not well-understood; for example, there exist certain classes of Lagrangian, for which the construction of the Hamiltonian is either ambiguous, or ill-defined \cite{roman2009multisymplectic}. Additionally, multisymplectic manifolds are not, in general, equipped with a well-defined bracket structure, and since the local constraint algorithm dispenses with such structures, our analysis of singular field theories will favour the \textit{Lagrangian} formalism, rather than the Hamiltonian picture employed for particles \cite{de1996geometrical,de2005pre}. In what follows, we provide a heavily abridged summary of the most pertinent ideas, referring to \cite{bell2025dynamicalsimilaritymultisymplecticfield} for further details.
\subsection{Lagrangian Field Theory}\label{Subsec:LagrangianFieldTheory}
In general, an $m$-dimensional manifold $\mathcal{M}$ is said to be multisymplectic if it admits a closed, 1-non-degenerate $k$-form $\Omega\in\Omega^k(\mathcal{M})$ (with $1<k\leqslant m$) \cite{roman2009multisymplectic}; the 1-non-degeneracy condition may be expressed locally as the requirement that for every $p\in \mathcal{M}$ and $X_p\in T_p\mathcal{M}$
\begin{equation*}
    \iota_{X_p}\Omega_p = 0 \quad\quad\implies \quad\quad X_p=0
\end{equation*}
If $\Omega$ is closed but 1-degenerate, we refer to the pair $(\mathcal{M},\Omega)$ as a pre-multisymplectic manifold. The field equations of a dynamical system are expressed geometrically in terms of multivector fields. A multivector field of degree $r$ on $\mathcal{M}$ is a section $\boldsymbol{X}\in \Gamma(\wedge^rT\mathcal{M})$ of the $r^{\textrm{th}}$ exterior power of the tangent bundle. We denote the space of all such multivector fields $\mathfrak{X}^r(\mathcal{M}):= \Gamma(\wedge^r\,T\mathcal{M})$, and declare some $\boldsymbol{X}\in\mathfrak{X}^r(\mathcal{M})$ to be \textit{locally decomposable} if, around $p\in\mathcal{M}$, there exists an open neighbourhood $\mathcal{U}_p\subset \mathcal{M}$, and vector fields $X_1,\,\cdots\,,X_r\in \mathfrak{X}^{\infty}(\mathcal{U}_p)$, such that $\boldsymbol{X}|_{\mathcal{U}_p} = X_1 \wedge\,\cdots\,\wedge X_r$. We say that an $m$-dimensional distribution $\mathcal{D}\subset T\mathcal{M}$ is \textit{locally associated} to a non-zero $\boldsymbol{X}\in \mathfrak{X}^m(\mathcal{M})$, if there exists some connected open set $\mathcal{V}\subset \mathcal{M}$, such that $\boldsymbol{X}|_{\mathcal{V}}$ is a section of $\wedge^m\mathcal{D}|_{\mathcal{V}}$; further, $\boldsymbol{X}$ is said to be \textit{integrable} if its locally associated distribution is integrable.\\

Consider a fibre bundle $\pi:E\rightarrow M$ over the $d$-dimensional spacetime manifold $M$; we shall suppose that $M$ is orientable, with volume form $\omega$. Local coordinates on $M$ are denoted $(x^{\mu})$, with $\mu=0,\,\cdots,d-1$, so that $\omega=dx^0\wedge \,\cdots \wedge dx^{d-1}:=d^dx$. The $(n+d)$-dimensional manifold $E$ is referred to as the covariant configuration space, and the first jet bundle $\kappa:J^1E\rightarrow E$ of sections of $\pi$ is the natural space upon which to introduce a Lagrangian density \cite{saunders1989geometry}; local adapted coordinates on $J^1E$ are given by $(x^{\mu},y^a,y^a_{\mu})$, with $a=1,\,\cdots,n$. Introducing the bundle projection $\widehat{\pi}:=\pi\circ\kappa:J^1E\rightarrow M$, the Lagrangian density $\mathcal{L}$ may be expressed as a $d$-form on $J^1E$
\begin{equation}\label{Eq:LagrangianDensity}
    \mathcal{L}(x^{\mu},y^a,y^a_{\mu}) = L(x^{\mu},y^a,y^a_{\mu})\,\widehat{\pi}^*\omega
\end{equation}  
in which $L:J^1E \rightarrow \mathbb{R}$ is referred to as the Lagrangian function, and $\widehat{\pi}^*\omega$ is the volume form on $M$, pulled back to $J^1E$. The Lagrangian function is then used to define the Cartan forms $\Theta_L\in\Omega^d(J^1E)$ and $\Omega_L\in\Omega^{d+1}(J^1E)$; in local bundle coordinates $(x^{\mu},y^a,y^a_{\mu})$, these are expressed as
\begin{equation}\label{Eq:CartanForms}
    \Theta_L = \frac{\partial L}{\partial y^a_{\mu}}\,dy^a\wedge d^{d-1}x_{\mu} - \left(\frac{\partial L}{\partial y^a_{\mu}}\,y^a_{\mu} - L\right)\,d^dx \quad\quad\quad\quad \Omega_L := -\,d\Theta_L
\end{equation}
where $d^{d-1}x_{\mu}:= \iota_{\scriptscriptstyle \partial_{\mu}} d^dx$. The pair $(J^1E,\Omega_L)$ defines a Lagrangian system, which is said to be \textit{regular} if $\Omega_L$ is multisymplectic, and \textit{singular} if it is pre-multisymplectic. Given a regular Lagrangian system $(J^1E,\Omega_L)$, the dynamical equations are derived from a variational principle \cite{roman2009multisymplectic}, which defines critical sections $\phi\in\Gamma(M,E)$; the canonical lifting $j^1\phi$ of these objects to $J^1E$ are then integral sections of an equivalence class of locally decomposable, $\widehat{\pi}$-transverse holonomic multivector fields $\{\boldsymbol{X}_L\}$, each of which satisfies
\begin{equation}
    \iota_{\boldsymbol{X}_L}\Omega_L=0
\end{equation}
The most general expression for a locally decomposable field $\boldsymbol{X}_L$ is
\begin{equation}\label{Eq:MultivectorFieldXL}
    \boldsymbol{X}_L = f\, \bigwedge\limits_{\mu=0}^{d-1} \left(\frac{\partial}{\partial x^{\mu}} + F^a_{\mu}\frac{\partial}{\partial y^a} + G^{a}_{\mu \nu}\frac{\partial}{\partial y^a_{\nu}} \right)
\end{equation}
for some non-zero $f\in C^{\infty}(J^1E)$. The $\widehat{\pi}$-transversality condition is most readily enforced by setting $\iota_{\boldsymbol{X}_L}(\widehat{\pi}^*\omega)=1$, which fixes the multiplicative function $f$ to unity. When $\boldsymbol{X}_L$ is holonomic, it is integrable, and the coefficient functions $F^a_{\mu}$ are simply $y^a_{\mu}$; if $\boldsymbol{X}_L$ has local coordinate expression (\ref{Eq:MultivectorFieldXL}), with $F^a_{\mu}=y^a_{\mu}$, but is \textit{not} integrable, it is referred to as semi-holonomic. The critical sections $\phi$ of the variational problem are such that their canonical lifting $j^1\phi$ satisfy the \textit{Euler-Lagrange field equations}
\begin{equation}\label{Eq:ELFieldEq}
    \frac{\partial}{\partial x^{\mu}}\left(\frac{\partial L}{\partial y^a_{\mu}}\circ j^1\phi\right) - \frac{\partial L}{\partial y^a}\circ j^1\phi=0
\end{equation}
\section{Pre-multicontact Systems}\label{Sec:PremulticontactSystems}
\subsection{Multicontact Lagrangian Field Theory}\label{Subsec:MulticontactLagrangianFieldTheory}
As described in the introduction, the excision of a superfluous scaling degree of freedom leads to a theory which is action-dependent, and thus has the interpretation of being non-conservative (The non-conservative/frictional character arises as a result of the fact that the contact manifolds upon which our action-dependent theories are defined do not respect a Liouville-like conservation of phase space volumes along the Hamiltonian flow). For field theories, this continues to be the case; now, however, the Lagrangian depends on the action \textit{density}. The fibred manifolds $\pi:E\rightarrow M$, and $\kappa:J^1E\rightarrow E$ (with $\textrm{dim}\,M=d$ and $\textrm{dim}\,E=n+d$) of the multisymplectic formalism continue to assume a prominent role in the description of frictional field theories; in particular, the Lagrangian density is now a $d$-form on the manifold
\begin{equation}\label{Eq:MulticontactConfigurationBundle}
    \mathcal{S} := J^1E \times_M \wedge^{d-1}T^*M \cong J^1E \times \mathbb{R}^d
\end{equation}
which is a bundle over both $E$, with projection $\tau:\mathcal{S}\rightarrow E$, and $M$, with $\beta=\pi\circ\tau:\mathcal{S}\rightarrow M$. Local coordinates on $\mathcal{S}$ are denoted $(x^{\mu},y^a,y^a_{\mu},s^{\mu})$, in which the quantities $s^{\mu}$ correspond to the action density mentioned above. As in (\ref{Eq:LagrangianDensity}), we write the Lagrangian density in terms of a local function
\begin{equation}\label{Eq:ContactLagrangian}
    \mathcal{L}(x^{\mu},y^a,y^a_{\mu},s^{\mu}) = L(x^{\mu},y^a,y^a_{\mu},s^{\mu})\,\beta^*\omega
\end{equation}
In contrast to the multisymplectic case, when the Lagrangian is singular, the $d$-form $\Theta_L$
\begin{equation}\label{Eq:LagrangianForm}
    \Theta_L = \left( ds^{\mu} - \frac{\partial L}{\partial y^a_{\mu}} dy^a\right) \,\wedge\, d^{d-1}x_{\mu} + \left(\frac{\partial L}{\partial y^a_{\mu}}\,y^a_{\mu} - L\right)\,d^dx
\end{equation}
does not necessarily define a pre-multicontact structure. It is therefore necessary that a number of additional criteria be met, which we now introduce. To facilitate this, we let $\Xi:=\beta^*\omega\in\Omega^d(\mathcal{S})$. While $\Theta_L$ endows $\mathcal{S}$ with the formal geometrical structure of a pre-multicontact manifold, $\Xi$ serves as a reference object. Given a regular distribution $\mathcal{D}\subset T\mathcal{S}$, we introduce the space of $r$-forms which annihilate all sections of $\mathcal{D}$
\begin{equation}\label{Eq:Ann}
    \textrm{Ann}^r(\mathcal{D}):= \{ \xi\in\Omega^r(\mathcal{S})\;|\; \iota_{X}\xi=0 \quad\textrm{for all}\;\,X\in\Gamma(\mathcal{D})\,\}
\end{equation}
The Reeb distribution associated with $(\mathcal{S},\Theta_L,\Xi)$ is then defined pointwise as
\begin{equation}\label{Eq:ReebDistribution}
    \mathcal{D}^R_p:=\{ X\in \textrm{ker}\,\Xi_p\;|\; \iota_Xd\Theta_L \in \textrm{Ann}^d_p(\textrm{ker}\,\Xi)\,\}
\end{equation}
Sections of $\mathcal{D}^R$ are the Reeb vector fields; we denote this space $\mathfrak{R}:=\Gamma(\mathcal{D}^R)$. In practice, we shall always work in coordinates such that $\Xi=dx^0\wedge\,\cdots\,\wedge dx^{d-1}$, so that $\textrm{ker}\,\Xi$ consists of those vector fields whose vertical components $V^{\mu}\partial_{x^{\mu}}$ are vanishing. The space $\textrm{Ann}^d(\textrm{ker}\,\Xi)$ then comprises all $d$-forms $\xi\in\Omega^d(\mathcal{S})$, for which $\iota_X\xi=0$ for any choice of vector field of the form
\begin{equation*}
    X = A^a\frac{\partial}{\partial y^a}+ B^a_{\mu}\frac{\partial}{\partial y^a_{\mu}} + C^{\mu}\frac{\partial}{\partial s^{\mu}}
\end{equation*}
In addition to the Reeb distribution, we also have the characteristic distribution $\mathcal{C}$, defined as the intersection $\mathcal{C}:=\textrm{ker}\,\Xi \cap \textrm{ker}\,\Theta_L\cap \textrm{ker}\,d\Theta_L$. The conditions under which the triple $(\mathcal{S},\Theta_L,\Xi)$ constitutes a pre-multicontact manifold are then that, for some $0<k\leqslant n(1+d)$
\begin{itemize}
    \item[$\star$] $\textrm{rank\;ker}\,\Xi=d+n+nd$
    \item[$\star$] $\textrm{rank}\,\mathcal{D}^R = d+k$
    \item[$\star$] $\textrm{rank}\,\mathcal{C}=k$
    \item[$\star$] $\textrm{Ann}^{d-1}(\textrm{ker}\,\Xi) = \{\iota_R\Theta_L\;|\; R\in\mathfrak{R}\,\}$
\end{itemize}
In practice, when working in adapted coordinates, the local Reeb fields are deduced from $\iota_{R_{\mu}}\Theta_L=d^{d-1}x_{\mu}$; however, note that this does not determine the $R_{\mu}$ uniquely, for it is possible to add to $R_{\mu}$ an element of the characteristic distribution, that is to say $\mathfrak{R}=\textrm{span}\hspace{0.4mm}(R_{\mu})+ \mathcal{C}$.\\

From $\Theta_L$, the $(d+1)$-form $\Omega_L$ is \textit{not} defined as in (\ref{Eq:CartanForms}) for multisymplectic systems, rather
\begin{equation}\label{Eq:ContactOmega}
    \Omega_L = d\Theta_L - \frac{\partial L}{\partial s^{\mu}}\,dx^{\mu} \wedge\,\Theta_L
\end{equation}
Given a (pre-)multicontact Lagrangian system $(\mathcal{S},\Theta_L)$, the equations of motion for sections $\Psi\in\Gamma(M,\mathcal{S})$ are given by
\begin{equation}\label{Eq:MulticontactEOM}
    \Psi^*\Theta_L = 0 \quad \quad \quad \textrm{and} \quad\quad\quad \Psi^*\iota_Z \Omega_L = 0 \quad \textrm{for all}\; Z\in \mathfrak{X}^{\infty}(\mathcal{S})
\end{equation}
We also have a similar pair equations for locally decomposable $\beta$-transverse multivector fields $\boldsymbol{X}_L\in\mathfrak{X}^d(\mathcal{S})$, which take the form
\begin{equation}\label{Eq:MulticontactEOM2}
    \iota_{\boldsymbol{X}_L}\Theta_L=0 \quad\quad\quad\quad\quad\iota_{\boldsymbol{X}_L}\Omega_L=0
\end{equation}
When $(\mathcal{S},\Theta_L)$ is multicontact, the multivector field solutions are integrable, with holonomic integral sections; in local coordinates, such a section may be expressed as
\begin{equation*}
    \Psi(x) = \left(x^{\mu},y^a(x), \frac{\partial y^a}{\partial x^{\mu}}\biggr|_x,s^{\mu}(x)\right)
\end{equation*}
with which the coordinate-free equations (\ref{Eq:MulticontactEOM}) become
\begin{equation}\label{Eq:HerglotzLagrangeEquations}
    \begin{split}
        \frac{\partial}{\partial x^{\mu}}\left(\frac{\partial L}{\partial y^a_{\mu}}\circ \Psi\right) = \left( \frac{\partial L}{\partial y^a}+\frac{\partial L}{\partial y^a_{\mu}}\frac{\partial L}{\partial s^{\mu}}\right) \circ \Psi\\
        \frac{\partial s^{\mu}}{\partial x^{\mu}} = L\circ \Psi
    \end{split}
\end{equation}
We refer to these expressions as the \textit{Herglotz-Lagrange field equations} \cite{Gaset_2024}. When $(\mathcal{S},\Theta_L)$ is pre-multicontact, if multivector field solutions of (\ref{Eq:MulticontactEOM2}) do exist, which is not guaranteed, they are generally not integrable. Consequently, the goal of the constraint algorithm, is to deduce the maximal submanifold $\mathcal{S}_f\hookrightarrow\mathcal{S}$ upon which holonomic multivector field solutions exist, and crucially, are tangent to $\mathcal{S}_f$.
\subsection{The Pre-Multicontact Constraint Algorithm}\label{Subsec:PreMulticontactAlgorithm}
Recall that the dynamical problem to be solved is the following: given a pre-multicontact Lagrangian system $(\mathcal{S},\Theta_L)$, we seek the maximal submanifold $S_f\hookrightarrow \mathcal{S}$, upon which there exist locally decomposable, $\beta$-transverse, holonomic solutions to (\ref{Eq:MulticontactEOM2}). We therefore begin by introducing the space of $d$-multivector fields which are solutions to the equations of motion, but not necessarily $\beta$-transverse or integrable
\begin{equation*}
    \textrm{ker}^d\,(\Theta_L,\Omega_L) := \{ \boldsymbol{X}_L\in \mathfrak{X}^d(\mathcal{S})\;|\; \iota_{\boldsymbol{X}_L}\Theta_L=0\quad\textrm{and}\quad\iota_{\boldsymbol{X}_L}\Omega_L = 0\,\}
\end{equation*}
Within this space, we then seek the subset of those multivector fields which are locally decomposable and $\beta$-transverse, denoting the resulting subspace $\textrm{ker}^d_{\beta}\,(\Theta_L,\Omega_L)$. In practice, as a first step, we shall simply assume that a solution $\boldsymbol{X}_L$ has the local coordinate decomposition
\begin{equation}\label{Eq:MultivectorField}
    \boldsymbol{X}_L = \bigwedge\limits_{\mu=0}^{d-1} \left(\frac{\partial}{\partial x^{\mu}} + F^a_{\mu}\frac{\partial}{\partial y^a} + G^{a}_{\mu \nu}\frac{\partial}{\partial y^a_{\nu}} + K_{\mu}^{\nu}\frac{\partial}{\partial s^{\nu}}\right)
\end{equation}
which is clearly an element of $\textrm{ker}^d_{\beta}\,(\Theta_L,\Omega_L)$. Upon substituting this into the dynamical equations $\iota_{\boldsymbol{X}_L}\Theta_L=0$ and $\iota_{\boldsymbol{X}_L}\Omega_L=0$, we must ensure that the resulting expressions do not contain inconsistencies. Supposing that we obtain a compatible set of equations, we now restrict $\textrm{ker}^d_{\beta}\,(\Theta_L,\Omega_L)$ to the subset of semi-holonomic solutions, which amounts to setting $F_{\mu}^a = y^a_{\mu}$ in the decomposition (\ref{Eq:MultivectorField}); in general, this step gives rise to further consistency conditions, and possibly to new constraints.\\

At every stage, we must ensure that the dynamics of the system remain confined to the constraint submanifold; thus, for each local constraint function $\Phi$, if $\boldsymbol{X}_L$ is expressed as the product $\boldsymbol{X}_L=X_0\wedge\cdots\wedge X_{d-1}$, we must demand that $\mathfrak{L}_{X_{\mu}}\Phi=0$ for $\mu=0,\,\cdots, d-1$. If these tangency conditions themselves give rise to additional constraints, such functions must again have vanishing Lie derivative along each of the $X_{\mu}$. The final step of the algorithm requires us to examine the integrability of our semi-holonomic solutions. Such an analysis is conducted by considering the constraints which arise as a result of imposing that $[X_{\mu},X_{\nu}]=0$ for $\mu,\nu=0,\,\cdots,d-1$. This leads to the final subspace $\textrm{ker}^d_{\scriptscriptstyle H}\,(\Theta_L,\Omega_L)$, consisting of locally decomposable, $\beta$-transverse, holonomic multivector fields, which, in general, will only exist on the submanifold $S_f\hookrightarrow \mathcal{S}$.
\section{Contact Reduction of Singular Field Theories}\label{Sec:ContactReductionField}
Following closely the framework developed in \cite{bell2025dynamicalsimilaritymultisymplecticfield}, we now present a field-theoretic generalisation of the ideas of section (\ref{Sec:ContactReduction1}). Let us consider the pre-multisymplectic system $(J^1E,\Omega_L)$, with corresponding Lagrangian function $L:J^1E\rightarrow \mathbb{R}$. From the Cartan $d$-form $\Theta_L$, we define
\begin{equation}\label{Eq:ThetaMu}
    \theta_L^{\mu}:=-\,\iota_{\partial_{d-1}}\cdots\; \iota_{\partial_0}\left(\Theta_L\wedge dx^{\mu}\right) = \frac{\partial L}{\partial y^a_{\mu}}dy^a
\end{equation}
The vector field $\Sigma\in\mathfrak{X}^{\infty}(J^1E)$ is said to constitute a scaling symmetry of degree $\Lambda$ if
\begin{equation}\label{Eq:DSConditions}
    \mathfrak{L}_{\Sigma}L=\Lambda L\quad\quad\quad\textrm{and}\quad\quad\quad \mathfrak{L}_{\Sigma}\theta_L^{\mu}=\theta_L^{\mu}\quad\textrm{for all}\;\mu=0,\,\cdots,d-1
\end{equation}
Consider a Herglotz (i.e multicontact) Lagrangian $L^H$, embedded within a multisymplectic manifold of one dimension higher through the expression
\begin{equation}\label{Eq:LandLH}
    L = e^{\rho}(L^H + \rho_{\mu}s^{\mu})
\end{equation}
in which $L^H$ depends upon the scalar fields $\phi^a$ ($a=1,\,\cdots ,k$), and their first derivatives $\phi^a_{\mu}$, together with the action density $s^{\mu}$, and the field $\rho$ is such that 
\begin{equation}\label{Eq:rho}
	\rho_{\mu} = -\, \frac{\partial L^H}{\partial s^{\mu}}
\end{equation}
It is straightforward to verify that the equations of motion derived from $L$ directly imply the Herglotz-Lagrange field equations for $L^H$, when the former is restricted to the subspace upon which $L^H$ is defined. In light of this, suppose that, having identified a scaling symmetry $\Sigma$ of our pre-multisymplectic system, we adopt coordinates on $J^1E$ in such a way so as to render this vector field of the form
\begin{equation}\label{Eq:Sigma}
    \Sigma=\xi\frac{\partial}{\partial \xi} + \xi_{\mu}\frac{\partial}{\partial \xi_{\mu}}
\end{equation}
In these coordinates, the Lagrangian function depends upon both $\xi$ and $\xi_{\mu}$, together with a set of unscaled field variables $\psi^a$, and their corresponding velocities $\psi^a_{\mu}$. Finally, we make the redefinition $\xi=e^{\rho/\Lambda}$, so that the scaling symmetry vector field is simply $\Sigma=\Lambda\hspace{0.2mm}\partial_{\rho}$, and the Lagrangian takes the form
\begin{equation}\label{Eq:L=e^rhof}
    L=e^{\rho}f(\rho_{\mu},\psi^a,\psi^a_{\mu})
\end{equation}
for some function $f$. The Euler-Lagrange equation for $\rho$ then implies that $f$ may be written as
\begin{align*}
    f = \rho_{\mu}\frac{\partial f}{\partial \rho_{\mu}} + \frac{\partial}{\partial x^{\mu}}\frac{\partial f}{\partial \rho_{\mu}}
\end{align*}
which, when compared to (\ref{Eq:LandLH}), and recalling that $\partial_{\mu}s^{\mu}=L^H$, suggests we should identify
\begin{equation}\label{Eq:ActionDensityandLH}
    s^{\mu} = \frac{\partial f}{\partial \rho_{\mu}} \quad\quad\quad\quad L^H = f - \rho_{\mu}s^{\mu}
\end{equation}
We must now take into account that $L$ was a singular Lagrangian. As such, the $d$-form $\Theta_{L^H}$ calculated from the symmetry-reduced function $L^H$ is not guaranteed to be pre-multicontact. Consequently, after making the `naïve' contact reduction, we must as an additional non-trivial step compute the characteristic and Reeb distributions. If these do not satisfy the conditions given in section (\ref{Sec:PremulticontactSystems}), the symmetry reduction procedure fails. For most physically interesting cases, this will not occur; however, the conditions must nevertheless be verified.\\

Having calculated $\Theta_{L^H}$, and confirmed that $\textrm{rank}\,\mathcal{D}^R = d+k$ and $\textrm{rank}\,\mathcal{C}=k$, for some $0<k\leqslant n(1+d)$, we have a pre-multicontact manifold $(\mathcal{S}_0,\Theta_{L^H})$, in which $\mathcal{S}_0$ denotes the reduced space obtained upon excising the scaling variable $\rho$. Geometrically, if the quantity $\xi$ in (\ref{Eq:Sigma}) corresponds to a single field variable on $E$, then upon writing the Lagrangian in the form (\ref{Eq:L=e^rhof}), $E$ separates into two connected pieces $E_{\pm}$, where each of $E_{\pm}$ is the product of a trivial bundle, and a codimension-1 subspace: $E_{\pm}\cong \widetilde{E}\times_M (M\times\mathbb{R}_{\pm})$. The change of variable $\xi\rightarrow e^{\rho/\Lambda}$ selects only the positive component of $\xi$; in order to cover the full range, we must also consider the change of variables $\xi\rightarrow -\,e^{\rho/\Lambda}$. Within each of $E_{\pm}$, it is precisely the trivial bundle that is eliminated by the contact reduction. Thus, provided we consider both components, so as not to lose any dynamical information, the reduced space is locally isomorphic to $J^1\widetilde{E}\times \mathbb{R}^d$, in which the codimension-1 subspace $\widetilde{E}$ is identified as a configuration space comprised of all field variables except $\rho$. Note that $\mathcal{S}_0$ is \textit{not} simply $J^1\widetilde{E}$ - such an outcome would be dimensionally inconsistent with the elimination of a single degree of freedom. Instead, the velocities $\rho_{\mu}$, which have not been eliminated, span a copy of $\mathbb{R}^d$ as action density components, thereby restoring the expected dimensional difference of $\textrm{dim}\,J^1E-\textrm{dim}\,\mathcal{S}_0=1$.\\

Supposing that $(\mathcal{S}_0,\Theta_{L^H})$ does constitute a pre-multicontact system, we calculate the $(d+1)$-form $\Omega_{L^H}$, and introduce a multivector field $\boldsymbol{X}_{L^H}\in\mathfrak{X}^d(\mathcal{S}_0)$, with local decomposition
\begin{equation*}
    \boldsymbol{X}_{L^H} = \bigwedge\limits_{\mu=0}^{d-1}\left(\frac{\partial}{\partial x^{\mu}} + F^a_{\mu}\frac{\partial}{\partial \psi^a} + G^a_{\mu\nu}\frac{\partial}{\partial \psi^a_{\nu}} + K^{\nu}_{\mu}\frac{\partial}{\partial s^{\nu}}\right)
\end{equation*}
in which we employ notation consistent with the decomposition (\ref{Eq:L=e^rhof}), denoting the unscaled fields $\psi^a$. Following this, we compute the constraints that arise as a result of setting $\iota_{\boldsymbol{X}_{L^H}}\Theta_{L^H}=0$ and $\iota_{\boldsymbol{X}_{L^H}}\Omega_{L^H}=0$, following the algorithmic procedure detailed in section (\ref{Subsec:PreMulticontactAlgorithm}). The final constraint submanifold $\mathcal{S}_f\hookrightarrow \mathcal{S}_0$ is the maximal space upon which holonomic multivector field solutions exist, and on this space, we faithfully reproduce the dynamics of the observables of the constrained pre-multisymplectic system.
\section{Effective Non-Abelian Gauge Theory}\label{Sec:Effective}
The appearance of scalar fields that directly couple to gauge curvature terms is a phenomenon which arises most notably in string-inspired models \cite{green2012superstring}. Indeed, the imprint of dilaton-like fields on low-energy effective actions has speculatively been regarded as a means to offer novel insight into problems such as colour confinement \cite{chabab2007confinement}. In this section, we present an example of a non-Abelian gauge theory coupled to a scalar field. Having implemented the reduction procedure, we discuss the physical implications of our results.
\subsection{Geometrical Setting}\label{Subsec:YMGeometrical}
We suppose that the spacetime manifold $M$ is equipped with a Lorentzian metric $g$ of signature $(+,-,-,\,\cdots\,)$, which is parametric, and non-variational \cite{gotay2004momentummapsclassicalrelativistic}, so that physically, our theory is defined on a curved background, but is not coupled to gravity. The appropriate covariant configuration space for this theory is 
\begin{equation}\label{Eq:Eg2CovariantConfiguration}
    \mathcal{E} = C(P) \times_M \left(M\times\mathbb{R}\right) \times_M \textrm{Sym}^{1,d-1}_2(M)
\end{equation}
Here, $P$ is the principal bundle over $M$, with structure group $G$, and $C(P)\rightarrow M$ refers to the bundle of connections on $P$, with $C(P)\cong J^1P/G$ \cite{LopezMC2001Tgot}. Finally, the space $\textrm{Sym}^{1,d-1}_2(M)$ denotes the set of symmetric covariant tensors of rank 2 and Lorentzian signature $(1,d-1)$ on $M$, and thus parameterises our choice of metric. Local coordinates on $\mathcal{E}$ are denoted $(x^{\mu},A^a_{\mu},\hspace{0.4mm}\phi\hspace{0.4mm};g_{\mu\nu})$, in which the semicolon separates the parametric and variational degrees of freedom, and the index $a$ runs from $a=1,\,\cdots,\textrm{dim}\,\mathcal{L}(G):=n$. The Lagrangian density is a $d$-form on the space 
\begin{equation}\label{Eq:Eg2CovariantVelocityPhase}
    \mathcal{Q}:=J^1\left(C(P) \times_M \left(M\times\mathbb{R}\right) \right) \times_M \textrm{Sym}^{1,d-1}_2(M)
\end{equation}
upon which we take local coordinates to be $(x^{\mu},A^a_{\mu},\hspace{0.4mm}\phi,A^a_{\mu,\nu},\hspace{0.4mm}\phi_{\mu}\hspace{0.4mm};g_{\mu\nu})$. The corresponding Lagrangian function $L:\mathcal{Q}\rightarrow \mathbb{R}$ is 
\begin{equation}\label{Eq:Eg2Lagrangian}
    L=\textrm{Tr}\left[-\frac{\phi^2}{2g^2}F_{\mu\nu}F^{\mu\nu} + 2J_{\mu}A^{\mu} + \frac{1}{2}g^{\mu\nu}\phi_{\mu}\phi_{\nu} - V(\phi)\right]
\end{equation}
where the trace is taken over the indices of $\mathcal{L}(G)$, and $F$ refers to the Lie algebra valued curvature 2-form. In what follows, we shall consider the dilaton to transform trivially under $G$, and adopt group-theoretic conventions more prevalent in the physics literature, taking the generators $T_a$ of the Lie algebra to be Hermitian, with $[T_a,T_b]=if_{ab}^{\;\;\;c}\,T_c$. In the fundamental representation, we adopt the following normalisation with respect to the trace
\begin{equation}\label{Eq:Eg2Trace}
    \textrm{Tr}\left(T_aT_b\right)=\frac{1}{2}\delta_{ab}
\end{equation}
We raise and lower Lie algebra indices with $\delta^{ab}$ and $\delta_{ab}$, writing $f_{abc}=\delta_{cd}\,f_{ab}^{\;\;\;d}$, for example (while we continue to raise and lower spacetime indices with $g^{\mu\nu}$ and $g_{\mu\nu}$). Returning to the Lagrangian (\ref{Eq:Eg2Lagrangian}), $J_{\mu}^a$ is a Lie algebra valued 1-form that couples to the gauge field, acting as a source term, and $V(\phi)$ is the non-perturbative dilaton potential. For simplicity of exposition, we shall take $V(\phi)$ to be a simple mass term $\frac{1}{2}m^2\phi^2$, and neglect the source. Thus, we have
\begin{equation}\label{Eq:Eg2Lagrangian2}
    L=-\,\frac{\phi^2}{4g^2}F^a_{\mu\nu}F_a^{\mu\nu} + \frac{1}{2}g^{\mu\nu}\phi_{\mu}\phi_{\nu} -\frac{1}{2}m^2\phi^2
\end{equation}
in which the curvature is expressed locally as $F_{\mu\nu}^a = A^a_{\nu,\mu} - A^a_{\mu,\nu} +f_{bc}^{\;\;\;a} A^b_{\mu}A^c_{\nu}$. The degeneracy of this Lagrangian arises as a result of the gauge symmetry; the Hessian matrix is, as expected, non-invertible. From (\ref{Eq:LagrangianForm}), we calculate the 1-forms $\theta_L^{\mu}$, finding that
\begin{equation}
    \theta^{\mu}_L:= -\,\iota_{\partial_{d-1}}\,\cdots\;\,\iota_{\partial_0}\left(\Theta_L\wedge dx^{\mu}\right) = -\,\frac{\phi^2}{g^2}F^{\mu\nu}_a\,dA^a_{\nu} + g^{\mu\nu}\phi_{\nu}\,d\phi
\end{equation}
From the structure of $\theta^{\mu}_L$, and the Lagrangian (\ref{Eq:Eg2Lagrangian2}), it is relatively clear that the vector field
\begin{equation*}
    \Sigma=\frac{1}{2}\left(\phi\frac{\partial}{\partial \phi} + \phi_{\mu}\frac{\partial}{\partial\phi_{\mu}}\right)
\end{equation*}
satisfies $\mathfrak{L}_{\Sigma}\theta_L^{\mu}=\theta^{\mu}_L$ and $\mathfrak{L}_{\Sigma}L=L$, with which we conclude that $\Sigma$ is a scaling symmetry of degree one. 
\subsection{Contact Reduction}\label{Subsec:Eg2ContactReduction}
Upon making the change of variables $\phi\mapsto e^{\rho/2}$, the scaling symmetry vector field $\Sigma$ is now simply $\partial_{\rho}$, and the Lagrangia reads
\begin{equation}\label{Eq:Eg2Lagrangian3}
    \begin{split}
        L = -\,\frac{e^{\rho}}{4g^2} F_{\mu\nu}^aF^{\mu\nu}_a +e^{\rho}\left(\frac{1}{8}g^{\mu\nu}\rho_{\mu}\rho_{\nu} - \frac{1}{2}m^2\right)
    \end{split}
\end{equation}
Recall that when the Lagrangian is expressed in the form $L=e^{\rho}f(\rho_{\mu},A_{\mu},A_{\mu,\nu})$, the action density is found from $s^{\mu}=\partial f/\partial \rho_{\mu}$, whilst the Herglotz Lagrangian takes the form $L^H=f-\rho_{\mu}s^{\mu}$; for the former, we find 
\begin{equation}\label{Eq:Eg2ActionDensity}
    s^{\mu}=\frac{\partial f}{\partial \rho_{\mu}} = \frac{1}{4}g^{\mu\nu}\rho_{\nu}\quad\quad\implies\quad\quad \rho_{\mu} = 4g_{\mu\nu}s^{\nu}
\end{equation}
while the Herglotz Lagrangian for this system is
\begin{equation}\label{Eq:Eg2HerglotzLag}
    L^H = -\,\frac{1}{4g^2}F_{\mu\nu}^aF^{\mu\nu}_a - \frac{1}{2}m^2 - 2g_{\mu\nu}s^{\mu}s^{\nu}
\end{equation}
Geometrically, referring to the decomposition (\ref{Eq:Eg2CovariantConfiguration}), we see that the excision of the dilaton corresponds to the removal of the trivial bundle $M\times\mathbb{R}\rightarrow M$, defining a reduced covariant configuration space $\mathcal{E}_{\textrm{red}}$. More formally, since $\phi$ is a global scaling function, the symplectification $\widetilde{\mathcal{E}}_{\textrm{red}}$ is a trivial $\mathbb{R}^{\times}$-bundle over $M$, composed of two connected components $\widetilde{\mathcal{E}}_{\textrm{red}}^{\pm}$, which, with the change of variables $\phi\rightarrow e^{\rho/2}$, correspond to symplectification via $\pm e^{\rho/2}$. Provided that both components are considered, we may somewhat informally write $\mathcal{E}_{\textrm{red}}\cong C(P)\times_M \textrm{Sym}^{1,d-1}_2(M)$. The Herglotz Lagrangian is defined on the space
\begin{equation}\label{Eq:Eg2ReducedQ}
    \mathcal{Q}_{\textrm{red}}= \left(J^1C(P)\times_M \textrm{Sym}_2^{1,d-1}(M)\right) \times\mathbb{R}^d
\end{equation}
where the additional factor of $\mathbb{R}^d$ comes from the $d$ velocities $\rho_{\mu}$, which have assumed the role of the action density $s^{\mu}$. The task is then to deduce whether the $d$-form $\Theta_{L^H}$, given by
\begin{equation}\label{Eq:Eg2HerglotzTheta}
    \Theta_{L^H} = \left(ds^{\mu} + \frac{1}{g^2}F^{\mu\nu}_a\,dA^a_{\nu}\right)\wedge d^{d-1}x_{\mu} + \left(-\,\frac{1}{g^2}F^{\mu\nu}_aA_{\nu,\mu}^a + \frac{1}{4g^2}F^{\mu\nu}_aF^a_{\mu\nu}+\frac{1}{2}m^2+2g_{\mu\nu}s^{\mu}s^{\nu}\right)d^dx
\end{equation}
endows $\mathcal{Q}_{\textrm{red}}$ with a pre-multicontact structure. Noting that $\iota_{\partial_{s^{\mu}}} \Theta_{L^H}= d^{d-1}x_{\mu}$, and that the coordinates $A^a_{\nu,\mu}$ only appear in the antisymmetric combinations $A^a_{\nu,\mu}-A^a_{\mu,\nu}$, it is straightforward to deduce that the characteristic and Reeb distributions are given by
\begin{equation}
    \mathcal{C} = \;\biggr\langle \frac{\partial}{\partial A^a_{\nu,\mu}}+\frac{\partial}{\partial A_{\mu,\nu}^a}\biggr\rangle \quad\quad\quad\quad \mathcal{D}^R = \;\biggr\langle \frac{\partial}{\partial A^a_{\nu,\mu}}+\frac{\partial}{\partial A_{\mu,\nu}^a}, \frac{\partial}{\partial s^{\mu}} \biggr\rangle
\end{equation}
for $\mu,\nu=0,\,\cdots,d-1$ and $a=1,\,\cdots, \textrm{dim}\,\mathcal{L}(G)=n$. The ranks of these distributions are
\begin{equation*}
    \textrm{rank}\,\mathcal{C}= n\,\frac{d(d+1)}{2} \quad\quad\quad \textrm{rank}\,\mathcal{D}^R = n\,\frac{d(d+1)}{2} + d
\end{equation*}
which, referring to the discussion of section (\ref{Subsec:MulticontactLagrangianFieldTheory}), are consistent with a pre-multicontact distribution with $k=d$. Consequently, we conclude that $\Theta_{L^H}$ does in fact define a pre-multicontact structure on $\mathcal{Q}_{\textrm{red}}$, and so may proceed with the constraint analysis; for the purpose of constructing multivector field solutions, we introduce the map $\beta:\mathcal{Q}_{\textrm{red}}\rightarrow M$, in-keeping with the notation of section (\ref{Subsec:PreMulticontactAlgorithm}).
\subsection{The Constraint Algorithm}\label{Subsec:Eg2ConstraintAlgorithm}
The constraint procedure commences supposing that a locally decomposable, $\beta$-transverse solution $\boldsymbol{X}\in \textrm{ker}^d_{\beta}\,(\Theta_{L^H},\Omega_{L^H})$  to the equations of motion exists, with
\begin{equation}\label{Eq:Eg2MultivectorField}
    \boldsymbol{X} = \bigwedge\limits_{\mu=0}^{d-1} \left(\frac{\partial}{\partial x^{\mu}} + C^a_{\mu\nu}\frac{\partial}{\partial A^a_{\nu}} + G^a_{\mu \nu\kappa}\frac{\partial}{\partial A^a_{\kappa,\nu}} + K_{\mu}^{\nu}\frac{\partial}{\partial s^{\nu}}\right) := \bigwedge\limits_{\mu=0}^{d-1} V_{\mu}
\end{equation}
The contraction of $\boldsymbol{X}$ with the pre-multicontact form $\Theta_{L^H}$ provides a single equation, as the degree of the multivector field coincides with that of the differential form. Imposing that $\iota_{\boldsymbol{X}}\Theta_{L^H}=0$, we find that
\begin{equation}\label{Eq:Eg2EOMTheta}
    K^{\mu}_{\mu} = \frac{1}{g^2} \left( A^a_{\nu,\mu}-C^a_{\mu\nu}\right) F^{\mu\nu}_a  + \left(-\,\frac{1}{4g^2}F^{\mu\nu}_aF^a_{\mu\nu}-\frac{1}{2}m^2-2g_{\mu\nu}s^{\mu}s^{\nu}\right)
\end{equation}
Note that the final quantity on the RHS is precisely $L^H$, and the bracket $\left( A^a_{\nu,\mu}-C^a_{\mu\nu}\right)$ vanishes upon imposing semi-holonomy. This expression therefore provides the vector-field analogue of the Herglotz equation $\partial_\mu s^\mu=L^H$ (c.f (\ref{Eq:HerglotzLagrangeEquations})). From (\ref{Eq:Eg2HerglotzTheta}), we compute that
\begin{equation}\label{Eq:Eg2HerglotzOmega}
    \begin{split}
        \Omega_{L^H} = \frac{1}{g^2}dF^{\mu\nu}_a\wedge \, dA^a_{\nu} \wedge d^{d-1}x_{\mu}+ \left(\frac{1}{2g^2}\left(F_a^{\mu\nu}dF_{\mu\nu}^a - 2F_a^{\mu\nu}\,dA^a_{\nu,\mu} - 2A^a_{\nu,\mu}\,dF^{\mu\nu}_a\right)\right.\\
        \left. - \frac{4}{g^2} g_{\mu\lambda}s^{\lambda}F^{\mu\nu}_a\,dA^a_{\nu}\right)\wedge d^dx
    \end{split}
\end{equation}
The contraction $\iota_{\boldsymbol{X}}\Omega_{L^H}$ then produces a pair of expressions, both of which must vanish separately
\begin{align}
    0 &= (A^a_{\nu,\mu}-A^a_{\mu,\nu})-(C^a_{\mu\nu}-C^a_{\nu\mu})\label{Eq:Eg2A}\\
    0 &= g^{\mu\rho}g^{\nu\sigma}\left(G_{\mu\rho\sigma}^a-G^a_{\mu\sigma\rho}\right) + f_{bc}^{\;\;\;a}g^{\nu\rho}A^{b\mu}\left(C^c_{\mu\rho}-C^c_{\rho\mu}\right) + f_{bc}^{\;\;\;a}\left(g^{\nu\rho}A^{b\mu} - g^{\mu\rho}A^{b\nu}\right)C^c_{\mu\rho}\label{Eq:Eg2B}\\
    &\quad+ f_{bc}^{\;\;\;e}f_{de}^{\;\;\;a}A^{b\mu}A^{c\nu}A^d_{\mu} + 4g_{\mu\lambda}s^{\lambda}F^{a\mu\nu}\notag
\end{align}
The next stage of the algorithm requires that we examine the effects of imposing semi-holonomy; for the multivector field (\ref{Eq:Eg2MultivectorField}), this implies we should set $C^a_{\mu\nu}=A^a_{\nu,\mu}$. With this, (\ref{Eq:Eg2A}) is rendered trivial, and (\ref{Eq:Eg2EOMTheta}) reduces to $K^{\mu}_{\mu}=L^H$, as expected. The first term of (\ref{Eq:Eg2B}) is then combined with $f_{bc}^{\;\;\;a}\left(g^{\nu\rho}A^{b\mu} - g^{\mu\rho}A^{b\nu}\right)C^c_{\mu\rho}$ to produce an object we shall denote $g^{\mu\rho}g^{\nu\sigma}F^a_{\rho\sigma,\mu}$. Finally, a number of the remaining terms vanish as a result of antisymmetry, with the result that (\ref{Eq:Eg2B}) becomes
\begin{equation}\label{Eq:Eg2EOM2}
    0 = g^{\mu\rho}g^{\nu\sigma}F^a_{\rho\sigma,\mu} + f_{bc}^{\;\;\;a}g^{\nu\rho}A^{b\mu}F^c_{\mu\rho} +4g_{\mu\lambda}s^{\lambda}F^{a\mu\nu}
\end{equation}
In the above, care has been taken to ensure that our notation for the velocity coordinates is merely reminiscent of a partial derivative with respect to the spacetime coordinates. In particular, until we have examined the integrability of our system, $A^a_{\nu,\mu}\neq \partial_{\mu}A^a_{\nu}$ and $G^a_{\mu\nu\sigma}\neq \partial_{\mu}\partial_{\nu} A^a_{\sigma}$. However, upon imposing the necessary conditions for $\textrm{span}\,(V_{\mu})$ to define an involutive distribution, we may affirm that the multivector field $\boldsymbol{X}$ possesses holonomic integral sections $\psi:M\rightarrow \mathcal{Q}_{\textrm{red}}$, whose local coordinate expression is
\begin{equation}\label{Eq:Eg2IntegralSection}
    \psi(x) = \left(x^{\mu},A^a_{\mu}(x),\frac{\partial A^a_{\mu}}{\partial x^{\nu}}(x),s^{\mu}(x)\right)
\end{equation}
Only after introducing these integral sections may we identify $A^a_{\nu,\mu}$ with the familiar $\partial_{\mu}A^a_{\nu}$. In light of this, we now turn to an analysis of the integrability of $\boldsymbol{X}$; the distribution formed by the $V_{\mu}$ will be involutive if, and only if it is closed under the Lie bracket; however, since partial derivatives with respect to the spacetime coordinates commute, involutivity is only guaranteed if $[V_{\mu},V_{\nu}]=0$ for all $\mu,\nu=0,\,\cdots,d-1$. Thus, computing the Lie bracket, and demanding that the result be zero, we find that
\begin{equation}\label{Eq:Eg2Integrability}
    G^a_{\nu\mu\sigma} - G^a_{\mu\nu\sigma} = A^a_{\sigma,\mu\nu}-A^a_{\sigma,\nu\mu}
\end{equation}
This condition produces no inconsistencies for the coefficients of the multivector field, and so (\ref{Eq:Eg2EOMTheta}) and (\ref{Eq:Eg2EOM2}) may be reexpressed as
\begin{equation}\label{Eq:Eg2FinalEoM}
    \begin{split}
        \partial_{\mu}s^{\mu} = L^H\\
        \partial_{\mu}F^{a\mu\nu} + f_{bc}^{\;\;\;a}A^b_{\mu}F^{c\mu\nu} + 4g_{\mu\lambda}s^{\lambda}F^{a\mu\nu}=0
    \end{split}
\end{equation}
The first two terms of the second expression form the familiar covariant derivative $(\mathcal{D}_{\mu}F^{\mu\nu})^a$, whilst the final term is a common feature of contact-reduced theories which we now explain. The scalar field $\phi$ is, in the original theory, the generator of a transformation to which the equations of motion of the remaining degrees of freedom are entirely insensitive. An observer internal to this system would be unable to detect rescalings of $\phi$ via measurement of the quantities to which they have ontological access. This is precisely the content of the discussion of the introduction, regarding the inaccessibility of changes in any degree of freedom without which the algebra of dynamical observables still closes. While $\phi$ is a superfluous degree of freedom, it still has a mathematical status within the original Lagrangian, coupling to the gauge-kinetic term $\textrm{Tr}\,(F_{\mu\nu}F^{\mu\nu})$. When we pass to a description which makes no reference to the dilaton, our theory becomes action-dependent, and thus non-conservative in nature. This is made manifest by the final term on the LHS of (\ref{Eq:Eg2FinalEoM}).
\section{Dynamical Coupling Parameters}\label{Sec:DynamicalCouplings}
\subsection{Reduction of Coupled Hamiltonian Systems}\label{Subsec:ReductionofCoupled}
In the previous section, we encountered a situation in which the closure of the algebra of dynamical observables was insensitive to the presence of the scalar field $\phi$. This may seem somewhat surprising, particularly for theories in which dimensionful parameters are determined dynamically by the expectation values of scalar fields. Let us assume, for the present discussion, that the expectation value $\langle\phi\rangle$ determines some coupling $g$. The value of $g$ is not directly accessible, but must be deduced via comparison to a reference object that we call measuring apparatus\footnote{For the purposes of the present discussion, it suffices to consider `measuring apparatus' in an abstract sense.}. This is entirely analogous to how an object of length 1\,m is only known to be such because its size is in a 1:1 ratio with that of a previously-standardised instrument. Consequently, changes in $\langle\phi\rangle$ are accompanied by simultaneous rescalings of the sensitivity of our apparatus, with which we would like to ascertain the value of $g$. It follows that the object responsible for these empirically-inaccessible changes in $g$ (i.e the field $\phi$) is redundant, from the perspective of the physical dynamics. This perspective also suggests that, since parameter values are necessarily ascertained through use of measuring apparatus that cannot be decoupled from the system to be measured, a scaling symmetry might also act on these parameters. Until now, we have only considered the vector field generator to act on the coordinates of phase space. However, the above line of argumentation is suggestive that we might consider an enlarged space upon which scaling symmetries can act. At present, a discussion of these ideas has only been formulated within the Hamiltonian description, while in the current work, we have favoured the Lagrangian perspective. We shall therefore attempt to outline the relevant ideas of the contact Reduction of Hamiltonian systems, referring to section (VI) of \cite{bell2025dynamicalsimilaritymultisymplecticfield} for further details.\\

The Hamiltonian is a function $H:J^1E^*\rightarrow \mathbb{R}$, where $J^1E^*$ denotes the multiphase space, upon which local coordinates (adapted to our scaling symmetry discussion) read $\left(x^{\mu},\rho,\phi^a,p_{\rho}^{\mu},p_a^{\mu}\right)$. Here, the $\phi^a$ are unscaled fields, and $p^{\mu}_a$ their corresponding momenta; note that, in stark contrast to more traditional approaches to classical and quantum field theory, each field variable has $d$ momenta - one for each value of $\mu$. Upon making a Legendre transform of the Lagrangian $L=e^{\rho}f(\rho_{\mu},\phi^a,\phi^a_{\mu})$ of section (\ref{Sec:ContactReductionField}), we find that the scaling symmetry vector field corresponding to $\Sigma=\Lambda\hspace{0.3mm}\partial_{\rho}$ reads
\begin{equation}\label{Eq:D}
    D=\frac{\partial}{\partial \rho} + p^{\mu}_{\rho}\frac{\partial}{\partial p^{\mu}_{\rho}} + p^{\mu}_{a}\frac{\partial}{\partial p^{\mu}_{a}}\hspace{2.5cm}p^{\mu}_a\;\textrm{unscaled momenta}
\end{equation}
This satisfies $\mathfrak{L}_DH=H$ and $\mathfrak{L}_D\hspace{0.3mm}\omega^{\mu}_H=\omega^{\mu}_H$, in which $\omega^{\mu}_H=d\rho\wedge dp_{\rho}^{\mu} + d\phi^a\wedge dp_a^{\mu}$, reminiscent of the familiar symplectic form in local Darboux coordinates. The reduction is then carried out defining coordinates on the reduced space, together with a multicontact Hamiltonian function. Letting $\tau: J^1E^*\rightarrow \mathcal{E}_{\textrm{red}}$ denote the projection to the reduced space $\mathcal{E}_{\textrm{red}}$, we have
\begin{equation}
    \tau^*\Pi^{\mu}_a :=\frac{p^{\mu}_a}{e^{\rho}} \quad\quad\quad \tau^*s^{\mu}:=\frac{p_{\rho}^{\mu}}{e^{\rho}}\quad\quad\quad \tau^*H^c(\phi^a,\Pi^{\mu}_a,s^{\mu}) := \frac{H^c}{e^{\rho}}
\end{equation}
The equations of motion for such a Hamiltonian read
\begin{equation}\label{Eq:ContactHamiltonEq}
    \frac{\partial \phi^a}{\partial x^{\mu}} = \frac{\partial H^c}{\partial p^{\mu}_a}\quad\quad \frac{\partial p^{\mu}_a}{\partial x^{\mu}} = -\left( \frac{\partial H^c}{\partial \phi^a} + p^{\mu}_a\frac{\partial H^c}{\partial s^{\mu}}\right) \quad\quad \frac{\partial s^{\mu}}{\partial x^{\mu}} = p^{\mu}_a\frac{\partial H^c}{\partial p^a_{\mu}} - H^c
\end{equation}
This procedure is applicable to a broad class of Hamiltonian systems; however, there often arise situations where the Hamiltonian comprises a number of pieces, each of which scales differently. Such a scenario significantly impedes our ability to reduce the theory, and the present goal is to provide a generalisation of the reduction framework, applicable to this particular type of theory. In order to make our discussion more concrete, we consider a Hamiltonian $H$ that is expressed as a sum $H=H_0 + g^i H_i$, for scalar couplings $g^i$ ($i=1,\,\cdots,N$, $g^i>0$). Here, following the definition of \ref{Eq:Deg}, $H_0$ is a degree-one Hamiltonian under (\ref{Eq:D}), and we assume that the remaining $H_i$ are of degree $\Lambda_i\neq 1$. It is the varied scaling behaviour of the $H_i$ that obstructs the standard contact reduction procedure.\\

Let us temporarily limit ourselves to a single coupling, so that $H=H_0+g H_1$, with $\mathfrak{L}_{D}H_0=H_0$ and $\mathfrak{L}_{D}H_1=\Lambda_1H_1$, for $\Lambda_1\neq 1$. To make the contact reduction, we extend the configuration space from $E$ to $E\times\mathbb{R}$, introducing the coordinate $X$. On $J^1\left(E\times\mathbb{R}\right)^*$, we acquire $d$ new momenta $P_X^{\mu}$, so that our coordinates become $(x^{\mu},\rho,\phi^a,X,p_{\rho}^{\mu},p_a^{\mu},P_X^{\mu})$.\\

The idea is now to assign the scalar coupling $g$ to some power of the Lorentz-invariant contraction $\eta_{\mu\nu}P_X^{\mu}P_X^{\nu}$. In particular, since $H_1$ is of degree $\Lambda_1$, we make the replacement
\begin{equation}\label{Eq:Dynamicalg}
    g\;\longrightarrow\; \left(\varepsilon\eta_{\mu\nu}P_X^{\mu}P_X^{\nu}\right)^\frac{1-\Lambda_1}{2}
\end{equation}
in which $\varepsilon=\pm$, as befits the signature of $\eta_{\mu\nu}$. On the extended space, we introduce $\bar{D}:=D+ P_X^{\mu}\partial_{P_X^{\mu}}$, and $\bar{\omega}^{\mu}_H:=\omega^{\mu}_H+dX\wedge dP_X^{\mu}$, writing the Hamiltonian as
\begin{equation}\label{Eq:HBar}
    \bar{H}=H_0 + \left(\varepsilon \eta_{\mu\nu}P_X^{\mu}P_X^{\nu}\right)^\frac{1-\Lambda_1}{2}H_1
\end{equation}
This construction guarantees that $\mathfrak{L}_{\bar{D}}\bar{H}=\bar{H}$, and $\mathfrak{L}_{\bar{\Sigma}}\bar{\omega}_H^{\mu}=\bar{\omega}_H^{\mu}$, and so we may make a standard degree-one symmetry reduction on $J^1(E\times\mathbb{R})^*$. Since $\bar{H}$ is independent of $X$, the field equations (\ref{Eq:ContactHamiltonEq}) tell us that $\partial_{\mu} P_X^{\mu}=0$. This is a divergence-free condition, and does \textit{not} imply that the $P_X^{\mu}$ are constant. We therefore introduce the following
\begin{equation}\label{Eq:Beta}
    \chi_{\mu}^{\;\;\nu}(x):=\partial_{\mu}P^{\nu}_X(x)
\end{equation}
where we have been deliberately insistent with the $x$ dependence. This is necessary because $P_X^{\mu}$ are, in general, coordinates on $J^1(E\times \mathbb{R})^*$, and so have no $x$ dependence. It is only on-shell, when we evaluate the equations of motion along some local section $\psi(x) = \left(x^{\mu},\rho(x),\phi^a(x),X(x),p_{\rho}^{\mu}(x), p^{\mu}_a(x),P_X^{\mu}(x)\right)$, that the $P_X^{\mu}$ acquire spacetime coordinate dependence. By introducing $P_X^\mu$ in place of the constant coupling $g$, we have effectively extended our theory, allowing for configurations not compatible with the original formulation. This extension must be executed with some degree of care, however. In particular, the extended Hamiltonian $\bar{H}$ is only well-defined when the quantity $P_X^2 := \eta_{\mu\nu} P_X^\mu P_X^\nu$ is greater than zero. Consequently, we restrict attention to a single connected component of the extended space, which we take to be $\mathcal{P}_+:=\{P_X^2 > 0\}$, excluding the singular locus $P_X^2 = 0$, and $\mathcal{P}_-:=\{P_X^2<0\}$, where the construction breaks down. Moreover, $P_X^\mu$ appears in the extended symmetry generator $\bar{D}$ through the combination $P_X^\mu \partial_{P_X^\mu}$, so that its action is given by scalings $P_X^\mu \to \lambda P_X^\mu$, with $\lambda \in \mathbb{R}_+$. It follows that the domain $\mathcal{P}_+$ is preserved under the flow of $\bar{D}$, ensuring consistency of the reduction within this region.\\

Having extended the dynamical content of our theory, the function $\chi_{\mu}^{\;\;\nu}(x)$ is precisely a measure of the departure from the dynamics of the constant momentum sector. Clearly $\chi_{\mu}^{\;\;\nu}(x)=0$ is a particular case within a larger space of possibilities. By working with the larger space of theories, we have introduced sufficient dynamical structure to make the contact reduction, but crucially, we retain the $\chi_{\mu}^{\;\;\nu}(x)=0$ case as a perfectly well-defined special instance, to which we are free to restrict ourselves post-reduction.\\

An additional observation we must make is that the contraction $\eta_{\mu\nu}P_X^{\mu}P_X^{\nu}$ is well-behaved only because we are in flat Minkowski space, and $\partial_{\lambda}\eta_{\mu\nu}=0$. When we pass to a general curved background, there arise two possible scenarios. On the one hand, we could treat the metric $g_{\mu\nu}$ parametrically, as in the Yang-Mills example. Alternatively, for a theory coupled to gravity, $g_{\mu\nu}$ is a dynamical variable that features in the variational calculation. In both cases, we must covariantise our expressions, replacing $\partial \rightarrow \nabla$. Thus, $\chi_{\mu}^{\;\;\nu}(x)=\nabla_{\mu}P_X^{\nu}(x)$, and then $\nabla_{\lambda}\left(g_{\mu\nu}P_X^{\mu}P_X^{\nu}\right)=0$.\\

All barred quantities, such as the Hamiltonian $\bar{H}$ and the modified scaling field $\bar{\Sigma}$, are defined on $J^1\left(E\times\mathbb{R}\right)^*$ (restricted, as we have stated above, to $\mathcal{P}_+$), and this is the space upon which the reduction is realised. From the extended configuration space $E\times\mathbb{R}$, we eliminate $\rho$, but the momenta $p^{\mu}_{\rho}$ remain, and become the components $s^{\mu}$ of the action density. Let $\mathcal{S}$ denote the space obtained from the reduction of the extended $J^1\left(E\times\mathbb{R}\right)^*$, and introduce the submersion $\bar{\tau}:J^1\left(E\times\mathbb{R}\right)^*\rightarrow \mathcal{S}$. Define reduced-space coordinates as
\begin{equation}\label{Eq:ReducedCoords}
    \bar{\tau}^*s^{\mu}:=\frac{p_{\rho}^{\mu}}{e^{\rho}} \quad\quad\quad\quad \bar{\tau}^*\Pi_a^{\mu}:=\frac{p_{a}^{\mu}}{e^{\rho}} \quad\quad\quad\quad \bar{\tau}^*\Pi_X^{\mu}:=\frac{P_{X}^{\mu}}{e^{\rho}}
\end{equation}
with all field coordinates unchanged: $\bar{\tau}^*\phi^a=\phi^a$. The contact Hamiltonian we write as $\bar{\tau}^*\bar{H}^c:=e^{-\rho}\bar{H}$, where the bar on $\bar{H}^c$ serves to remind us that we are still on the extended reduced space, and have not yet reduced the number of degrees of freedom. The equations of motion are then deduced from $\bar{H}^c$ precisely as in (\ref{Eq:ContactHamiltonEq}). Temporarily dropping submersions, we note that the vanishing of $\chi_{\mu}^{\;\;\nu}(x)$ may be expressed in reduced-space coordinates as follows
\begin{align*}
    \chi_{\mu}^{\;\;\nu}(x) = \partial_{\mu}\left(e^{\rho}\Pi_X^{\nu}\right)= e^{\rho}\left(\rho_{\mu}\Pi_X^{\nu}+ \partial_{\mu}\Pi_X^{\nu}\right)= e^{\rho}\underbrace{\left(\frac{\partial \bar{H}^c}{\partial s^{\mu}}\Pi_X^{\nu}+ \partial_{\mu}\Pi_X^{\nu}\right)}_{:=\,^c\chi_{\mu}^{\;\;\nu}}
\end{align*}
In the final equality, we have introduced the `contact $\chi$-function'. Clearly, we require that $^c\chi_{\mu}^{\;\;\nu}(x)=0$, since $e^{\rho}\neq0$. Note that, contracting $\mu$ and $\nu$, the vanishing of $^c\chi_{\mu}^{\;\;\mu}(x)$ is fully consistent with the multicontact equations of motion derived from $\bar{H}^c$
\begin{equation*}
    \partial_{\mu}\Pi_X^{\mu}=-\left(\frac{\partial \bar{H}^c}{\partial X} + \Pi_X^{\mu}\frac{\partial \bar{H}^c}{\partial s^{\mu}}\right) \quad\quad\implies \quad\quad \partial_{\mu}\Pi_X^{\mu}=-  \Pi_X^{\mu}\frac{\partial \bar{H}^c}{\partial s^{\mu}}
\end{equation*}
which is precisely the statement that $^c\chi_{\mu}^{\;\;\mu}(x)=0$.\\

The reader will have noticed that we have performed a contact reduction on the extended space $J^1(E\times\mathbb{R})^*$. While the number of degrees of freedom has certainly decreased with respect to this space, the same is not true of the original multiphase space. That is to say, at present, the reduced space $\mathcal{S}$ is such that $\textrm{dim}\,\mathcal{S}>\textrm{dim}\,J^1E^*$. It therefore remains to be justified that we have in fact successfully reduced the number of degrees of freedom of the original system. In order to be more concrete, the dimension of the original multiphase space is $\textrm{dim}\,J^1E^*=d+n(1+d)$; for each coupling parameter $g_i$, we introduce one field variable $X_i$, which is accompanied by $d$ momenta $P_{X_i}^{\mu}$. Consequently, if we have $N$ coupling parameters, the dimension of the extended space $J^1(E\times\mathbb{R}^N)^*$ is $d+n(1+d)+N(1+d)$. The contact reduction on the extended space then reduces this by one; however, we still have $\textrm{dim}\,\mathcal{S}-\textrm{dim}\,J^1E^* = N(1+d)-1$.\\

The resolution of this apparent discrepancy lies in the fact that the additional degrees of freedom introduced to make the reduction are largely redundant. In particular, $\mathcal{S}$ carries a natural $\mathbb{R}^N$-action generated by the vector fields $V_i:=\bar{\tau}_*\partial_{X_i}$, corresponding to translations in the $X_i$. As well as this, while the contact momenta $\Pi^{\mu}_{X_i}$ themselves are not constant, we may construct $Nd$ independent invariant quantities by taking ratios of the components of the momenta. By restricting to a regular level-set of these invariants, and quotienting out by the (free) $\mathbb{R}^N$-action, we remove precisely the excess $N(1+d)$ degrees of freedom. This gives a space of dimension $\textrm{dim}\,J^1E^*-1$, as expected for a contact reduction.
\subsection{A Simple Example}\label{Subsec:Eg1}
In order to illustrate how this formalism is implemented in practice, we shall consider the example of two scalar fields $\phi$ and $\psi$, with a quartic interaction term. We take the Lagrangian function to be
\begin{equation}\label{Eq:Eg1Lag1}
    L = \frac{1}{2}\eta^{\mu\nu}\phi_{\mu}\phi_{\nu}+ \frac{1}{2}\eta^{\mu\nu}\psi_{\mu}\psi_{\nu} -\frac{1}{2}m^2\left(\phi^2+\psi^2\right) - \frac{\lambda}{4!}\left(\phi^4-\psi^4\right)
\end{equation}
Introducing spherical polars $\phi=R\sin \theta$ and $\psi=R\cos\theta$, we restrict to $R>0$ for simplicity, which allows us to parameterise $R=e^{\rho/2}$, with which the Lagrangian becomes
\begin{equation}\label{Eq:Eg1Lag2}
    L=e^{\rho}\biggr[\eta^{\mu\nu}\left( \frac{1}{8}\rho_{\mu}\rho_{\nu} + \theta_{\mu}\theta_{\nu}\right) -\frac{1}{2}m^2 \biggr] + \frac{\lambda}{4!}e^{2\rho}\cos2\theta
\end{equation}
Note that the sign in front of the quartic interaction term has changed; this is due to the fact that we get a $(\sin^2\theta - \cos^2\theta)=-\,\cos2\theta$ term. It is then an easy exercise to calculate the momenta and pass to the following Hamiltonian
\begin{equation}\label{Eq:Eg1Ham1}
    H = e^{-\rho}\eta_{\mu\nu}\left(2p^{\mu}_{\rho}p^{\nu}_{\rho} + \frac{1}{2}p^{\mu}_{\theta}p^{\nu}_{\theta}\right) + \frac{1}{2}m^2e^{\rho} - \frac{\lambda}{4!}e^{2\rho}\cos2\theta
\end{equation}
From our discussion, it is clear we take
\begin{equation*}
    H_0:=e^{-\rho}\eta_{\mu\nu}\left(2p^{\mu}_{\rho}p^{\nu}_{\rho} + \frac{1}{2}p^{\mu}_{\theta}p^{\nu}_{\theta}\right) + \frac{1}{2}m^2e^{\rho}\quad\quad\quad H_1:=-\,\frac{1}{4!}e^{2\rho}\cos2\theta
\end{equation*}
so that $H=H_0+\lambda H_1$, with $\Lambda_1=2$ under the obvious scaling symmetry
\begin{equation}
    \Sigma=\frac{\partial}{\partial \rho} + p^{\mu}_{\rho}\frac{\partial}{\partial p^{\mu}_{\rho}} + p^{\mu}_{\theta}\frac{\partial}{\partial p^{\mu}_{\theta}}
\end{equation}
Thus, we introduce $(X,P_X^{\mu})$, and the exponent of the Lorentz-invariant contraction is $(1-\Lambda_1)/2=-1/2$. Restricting to the domain $\{P_X^2>0\}$, so that our expressions are real and well-defined, the Hamiltonian becomes
\begin{equation}
    \bar{H} = e^{-\rho}\eta_{\mu\nu}\left(2p^{\mu}_{\rho}p^{\nu}_{\rho} + \frac{1}{2}p^{\mu}_{\theta}p^{\nu}_{\theta}\right) + \frac{1}{2}m^2e^{\rho} - \frac{\left(\eta_{\lambda\sigma}P_X^{\lambda}P_X^{\sigma}\right)^{-1/2}}{4!}\,e^{2\rho}\cos2\theta
\end{equation}
The modified scaling vector field
\begin{equation}
    \bar{\Sigma}=\frac{\partial}{\partial \rho} + p^{\mu}_{\rho}\frac{\partial}{\partial p^{\mu}_{\rho}} + p^{\mu}_{\theta}\frac{\partial}{\partial p^{\mu}_{\theta}} + P_X^{\mu}\frac{\partial}{\partial P_X^{\mu}}
\end{equation}
is of degree one for $\bar{H}$, as expected. Taking coordinates on the reduced space exactly as in (\ref{Eq:ReducedCoords}), the contact Hamiltonian reads
\begin{equation}\label{Eq:Eg1ContactHam1}
    \bar{H}^c = \eta_{\mu\nu}\left(2s^{\mu}s^{\nu} + \frac{1}{2}\Pi^{\mu}_{\theta}\Pi^{\nu}_{\theta}\right) + \frac{1}{2}m^2 - \frac{\left(\eta_{\lambda\sigma}\Pi_X^{\lambda}\Pi_X^{\sigma}\right)^{-1/2}}{4!}\,\cos2\theta
\end{equation}
The equations of motion of the original multisymplectic Hamiltonian (\ref{Eq:Eg1Ham1}) are
\begin{align}\label{Eq:Eg2MultisimpEoM}
    \partial_{\mu}\rho &= 4e^{-\rho}\eta_{\mu\nu}p_{\rho}^{\nu} & \partial_{\mu}p_{\rho}^{\mu}&=e^{-\rho}\eta_{\mu\nu}\left(2p^{\mu}_{\rho}p^{\nu}_{\rho} + \frac{1}{2}p^{\mu}_{\theta}p^{\nu}_{\theta}\right)- \frac{1}{2}m^2e^{\rho} + \frac{2\lambda}{4!}e^{2\rho}\cos2\theta\\
    \partial_{\mu}\theta &= e^{-\rho}\eta_{\mu\nu}p_{\theta}^{\nu} & \partial_{\mu}p^{\mu}_{\theta} &= -\,\frac{2\lambda}{4!}e^{2\rho}\sin 2\theta\notag
\end{align}
These must be replicated in the reduced system, upon restricting to a suitable regular level-set of the invariants. The contact Hamiltonian equations that follow from (\ref{Eq:Eg1ContactHam1}) are
\begin{align}
    \partial_{\mu}\theta &= \eta_{\mu\nu}\Pi_{\theta}^{\nu} & \partial_{\mu}\Pi_{\theta}^{\mu}&= -\left(\frac{2\left(\eta_{\lambda\sigma}\Pi_X^{\lambda}\Pi_X^{\sigma}\right)^{-1/2}}{4!}\,\sin2\theta + 4\eta_{\mu\nu}s^{\nu}\Pi_{\theta}^{\mu}\right)\notag\\
    \partial_{\mu}X &= \frac{\left(\eta_{\lambda\sigma}\Pi_X^{\lambda}\Pi_X^{\sigma}\right)^{-3/2}}{4!}\,\frac{\eta_{\mu\nu}\Pi_X^{\nu}}{2}\cos2\theta & \partial_{\mu}\Pi^{\mu}_{X} &= -\,4\eta_{\mu\nu}\Pi_X^{\mu} s^\nu
\end{align}
\vspace{-0.6cm}
\begin{equation*}
    \partial_{\mu}s^{\mu} = \eta_{\mu\nu}\left(-\,2s^{\mu}s^{\nu} + \frac{1}{2}\Pi^{\mu}_{\theta}\Pi^{\nu}_{\theta}\right) - \frac{1}{2}m^2 + \frac{2\left(\eta_{\lambda\sigma}\Pi_X^{\lambda}\Pi_X^{\sigma}\right)^{-1/2}}{4!}\,\cos2\theta
\end{equation*}
It is then a relatively straightforward (albeit lengthy) exercise to use the definitions of the contact momenta $\Pi^{\mu}$ to show that the dynamics of the original multisymplectic system is captured by the contact system. This calculation is carried out after having restricted to a regular level-set of the invariants.
\section{Conclusions and Outlook}\label{Sec:Conclusions}
Many of the most mathematically rich models in contemporary theoretical physics are characterised by the presence of gauge symmetries, and are therefore described by singular Lagrangians. In this work, we have shown that a broad class of such theories — namely, those admitting global scaling symmetries — may be subjected to a contact reduction procedure analogous to that previously developed for regular systems. A key aspect of this construction is the consistent treatment of degeneracies: we have demonstrated that the associated constraint algorithm may be implemented either pre or post-reduction, with both procedures leading to equivalent results. This compatibility reflects the fact that the constraint structure is insensitive to the global scaling degree of freedom, and from a physical perspective, is unsurprising. Non-commutativity of the processes of contact reduction and phase space restriction would be in direct contradiction of the unobservability of changes in a variable that does not contribute to the closure of the algebra of the ontologically-accessible degrees of freedom.\\

The reduction of singular field theories was illustrated through the example of an effective non-Abelian gauge theory coupled to a dilaton-like scalar. It was observed that this field was redundant, and could therefore be excised from our description. The superfluous nature of this degree of freedom was not in contradiction with the notion that coupling parameters of a theory are often fixed by the expectation values of scalar fields. Instead, we argued that the strength of an interaction is only meaningful relative to some choice of scale, and that any operational procedure conducted with the aim of measuring the strength of such interactions will be insensitive to a rescaling of their values. This is a direct consequence of the fact that the apparatus used is not separable from the system to be measured.\\

The framework presented throughout provides a systematic approach to the identification and subsequent elimination of scaling degrees of freedom in singular (field) theories. This is particularly relevant in gravitational contexts, where it is known that, upon expressing the spacetime metric as a product of a conformal factor and a symmetric rank-two tensor of fixed determinant, the Einstein-Hilbert action possesses a scaling symmetry, corresponding precisely to changes in the conformal factor \cite{sloan2025dynamical}. Our results suggest that it is possible to eliminate this degree of freedom prior to implementing a constraint algorithm, thereby working directly with a reduced ontology. This raises the possibility of reformulating gravitational dynamics without explicit reference to scale. Such a description may prove highly advantageous, particularly in those regimes in which the conventional formulation becomes singular, and therefore ceases to be predictive.\\

Finally, more work is required to consolidate a fully satisfactory field-theoretic implementation of the promotion of coupling parameters to dynamical variables. Developing the prescription in the Lagrangian framework would provide a means to analyse a much broader class of singular theories within a unified geometric setting.
\section*{Acknowledgements}
We would like to thank the anonymous referee, whose valuable feedback contributed to the revision and improvement of this manuscript. DS also acknowledges generous support from the Foundational Questions Institute.
\bibliographystyle{unsrt}
\bibliography{Refs}

\begin{thebibliography}{10}

\bibitem{peskin2018introduction}
Michael~E Peskin.
\newblock {\em An Introduction to quantum field theory}.
\newblock CRC press, 2018.

\bibitem{palti2019swampland}
Eran Palti.
\newblock The swampland: introduction and review.
\newblock {\em Fortschritte der Physik}, 67(6):1900037, 2019.

\bibitem{leibniz1989discourse}
Gottfried~Wilhelm Leibniz.
\newblock Discourse on metaphysics: 1686.
\newblock In {\em Philosophical papers and letters}, pages 303--330. Springer, 1989.

\bibitem{sloan2018dynamical}
David Sloan.
\newblock Dynamical similarity.
\newblock {\em Physical Review D}, 97(12):123541, 2018.

\bibitem{carinena2013canonoid}
Jos{\'e}~F Cari{\~n}ena, Fernando Falceto, and Manuel~F Ranada.
\newblock Canonoid transformations and master symmetries.
\newblock {\em Journal of Geometric Mechanics}, 5(2), 2013.

\bibitem{sloan2021scale}
David Sloan.
\newblock Scale symmetry and friction.
\newblock {\em Symmetry}, 13(9):1639, 2021.

\bibitem{bell2025dynamicalsimilaritymultisymplecticfield}
Callum Bell and David Sloan.
\newblock Dynamical similarity in multisymplectic field theory, 2025.

\bibitem{Nester1}
Mark~J. Gotay, James~M. Nester, and George Hinds.
\newblock Presymplectic manifolds and the dirac-bergmann theory of constraints.
\newblock {\em Journal of Mathematical Physics}, 19(11):2388--2399, 11 1978.

\bibitem{bell2026constrained}
Callum Bell and David Sloan.
\newblock Constrained symplectic and contact hamiltonian systems: A review.
\newblock {\em arXiv preprint arXiv:2604.27940}, 2026.

\bibitem{roman2009multisymplectic}
Narciso Rom{\'a}n-Roy et~al.
\newblock Multisymplectic lagrangian and hamiltonian formalisms of classical field theories.
\newblock {\em SIGMA. Symmetry, Integrability and Geometry: Methods and Applications}, 5:100, 2009.

\bibitem{de2023multicontact}
Manuel de~Le{\'o}n, Jordi Gaset, Miguel~C Mu{\~n}oz-Lecanda, Xavier Rivas, and Narciso Rom{\'a}n-Roy.
\newblock Multicontact formulation for non-conservative field theories.
\newblock {\em Journal of Physics A: Mathematical and Theoretical}, 56(2):025201, 2023.

\bibitem{dick1997implications}
Rainer Dick.
\newblock Implications of a dilaton in gauge theory and cosmology.
\newblock {\em Fortschritte der Physik/Progress of Physics}, 45(7):537--587, 1997.

\bibitem{sloan2025dynamical}
David Sloan.
\newblock Dynamical similarity in field theories.
\newblock {\em Classical and Quantum Gravity}, 42(4):045001, 2025.

\bibitem{de2019contact}
Manuel de~Le{\'o}n and Manuel Lainz~Valc{\'a}zar.
\newblock Contact hamiltonian systems.
\newblock {\em Journal of Mathematical Physics}, 60(10), 2019.

\bibitem{de2020review}
Manuel de~Le{\'o}n and Manuel Lainz.
\newblock A review on contact hamiltonian and lagrangian systems.
\newblock {\em arXiv preprint arXiv:2011.05579}, 2020.

\bibitem{bravetti2019contact}
Alessandro Bravetti.
\newblock Contact geometry and thermodynamics.
\newblock {\em International Journal of Geometric Methods in Modern Physics}, 16(supp01):1940003, 2019.

\bibitem{ismael2021symmetry}
Jenann Ismael.
\newblock Symmetry and superfluous structure: Lessons from history and tempered enthusiasm.
\newblock In {\em The Routledge Companion to Philosophy of Physics}, pages 563--577. Routledge, 2021.

\bibitem{ismael2003symmetry}
Jenann Ismael and Bas~C Van~Fraassen.
\newblock Symmetry as a guide to superfluous theoretical structure.
\newblock {\em Symmetries in physics: Philosophical reflections}, pages 371--392, 2003.

\bibitem{sloan2019scalar}
David Sloan.
\newblock Scalar fields and the flrw singularity.
\newblock {\em Classical and Quantum Gravity}, 36(23):235004, 2019.

\bibitem{bravetti2023scaling}
Alessandro Bravetti, Connor Jackman, and David Sloan.
\newblock Scaling symmetries, contact reduction and poincar{\'e}’s dream.
\newblock {\em Journal of Physics A: Mathematical and Theoretical}, 56(43):435203, 2023.

\bibitem{cantrijn1999geometry}
Frans Cantrijn, Alberto Ibort, and Manuel De~Le{\'o}n.
\newblock On the geometry of multisymplectic manifolds.
\newblock {\em Journal of the Australian Mathematical Society}, 66(3):303--330, 1999.

\bibitem{forger2013multisymplectic}
Michael Forger and Leandro~G Gomes.
\newblock Multisymplectic and polysymplectic structures on fiber bundles.
\newblock {\em Reviews in Mathematical Physics}, 25(09):1350018, 2013.

\bibitem{de1996geometrical}
Manuel de~Le{\'o}n, Jes{\'u}s Marin-Solano, and Juan~C Marrero.
\newblock A geometrical approach to classical field theories: a constraint algorithm for singular theories.
\newblock In {\em New Developments in Differential Geometry: Proceedings of the Colloquium on Differential Geometry, Debrecen, Hungary, July 26--30, 1994}, pages 291--312. Springer, 1996.

\bibitem{de2005pre}
Manuel De~Le{\'o}n, Jes{\'u}s Mar{\'\i}n-Solano, Juan~Carlos Marrero, Miguel~C Munoz-Lecanda, and Narciso Rom{\'a}n-Roy.
\newblock Pre-multisymplectic constraint algorithm for field theories.
\newblock {\em International Journal of Geometric Methods in Modern Physics}, 2(05):839--871, 2005.

\bibitem{saunders1989geometry}
David~J Saunders.
\newblock {\em The geometry of jet bundles}, volume 142.
\newblock Cambridge University Press, 1989.

\bibitem{Gaset_2024}
Jordi Gaset, Manuel Lainz, Arnau Mas, and Xavier Rivas.
\newblock The herglotz variational principle for dissipative field theories.
\newblock {\em Geometric Mechanics}, 01(02):153–178, June 2024.

\bibitem{green2012superstring}
Michael~B Green, John~H Schwarz, and Edward Witten.
\newblock {\em Superstring theory: volume 2, loop amplitudes, anomalies and phenomenology}.
\newblock Cambridge university press, 2012.

\bibitem{chabab2007confinement}
Mohamed Chabab.
\newblock Confinement driven by scalar field in 4d non abelian gauge theories.
\newblock In {\em AIP Conference Proceedings}, volume 881, pages 238--245. American Institute of Physics, 2007.

\bibitem{gotay2004momentummapsclassicalrelativistic}
Mark~J. Gotay, James Isenberg, Jerrold~E. Marsden, and Richard Montgomery.
\newblock Momentum maps and classical relativistic fields. part i: Covariant field theory.
\newblock 2004.

\bibitem{LopezMC2001Tgot}
MC~Lopez and JM~Masque.
\newblock The geometry of the bundle of connections.
\newblock {\em Mathematische Zeitschrift}, 236(4):797--811, 2001.

\end{thebibliography}
\end{document}